\documentclass[api, phf,16pt,showpacs, reprint,onecolumn]{revtex4-1}

\usepackage{amssymb,amsfonts,latexsym,mathtools} %
\usepackage{graphicx,setspace}
\usepackage{graphicx} %
\usepackage{amsmath,euscript,mathrsfs}
\usepackage[hidelinks]{hyperref} 
\usepackage[usenames,dvipsnames,svgnames]{xcolor}

\hypersetup{
    colorlinks,
    linkcolor={blue!50!black},
    citecolor={blue!50!black},
    urlcolor={blue!50!black}
}

\graphicspath{{Figures/}}

\makeatletter
\def\blfootnote{\gdef\@thefnmark{}\@footnotetext}
\makeatother

\usepackage[all]{hypcap}

\usepackage{array}
\newcolumntype{C}[1]{>{\centering\arraybackslash}m{#1}}

\usepackage[super]{nth}

\usepackage{comment}

\DeclarePairedDelimiter{\abs}{\lvert}{\rvert}

\allowdisplaybreaks[1]

\newcommand{\nc}{\newcommand}
\nc{\numberthis}{\addtocounter{equation}{1}\tag{\theequation}}

\nc{\be}{\begin{equation}}
\nc{\ee}{\end{equation}}
\nc{\bes}{\begin{equation*}}
\nc{\ees}{\end{equation*}}
\nc{\eps}{\varepsilon}
\nc{\prt}{\partial}
\nc{\ds}{\displaystyle}

\nc{\dsp}[1]{^\mathrm{#1}}
\nc{\lb}{\left (}
\nc{\rb}{\right )}
\nc{\lset}{\left \{}
\nc{\rset}{\right \}}
\nc{\eqtext}[1]{\quad \text{#1} \quad}
\nc{\lsq}{\left [}
\nc{\rsq}{\right ]}

\nc{\half}{\frac{1}{2}}
\nc{\RA}{\quad \Rightarrow \quad}
\nc{\coshn}[2]{\mathop{\rm cosh}\nolimits ^{#1} \lb #2 \rb}
\nc{\sechn}[2]{\mathop{\rm sech}\nolimits ^{#1} \lb #2 \rb}
\nc{\tanhn}[2]{\mathop{\rm tanh}\nolimits ^{#1} \lb #2 \rb}
\nc{\arccosh}[1]{\mathop{\rm arccosh}\nolimits \lb #1 \rb}
\nc{\ob}[1]{\overbrace{#1}}
\nc{\ub}[1]{\underbrace{#1}}
\nc{\field}[1]{\mathbb{#1}}
\nc{\inflim}{_{-\infty}^{\infty}}

\nc{\dd}[1]{\; \mathrm{d} #1}
\nc{\diff}[2]{\frac{\mathrm{d} #1}{\mathrm{d} #2}}
\nc{\diffn}[3]{\dfrac{\mathrm{d}^{#1} #2}{\mathrm{d} #3^{#1}}}
\nc{\pdd}[1]{\; \partial #1}
\nc{\pdiff}[2]{\frac{\partial #1}{\partial #2}}
\nc{\pdiffn}[3]{\dfrac{\partial^{#1} #2}{\partial #3^{#1}}}

\nc{\nt}{\newtheorem}
\nc{\ntc}{\newtheorem*}

\nt{theorem}{Theorem}[section]
\nt{definition}{Definition}[section]
\nt{exa}{Example}[section]
\nt{lem}{Lemma}[section]
\nt{cor}{Corollary}[section]
\nt{pro}{Proposition}[section]

\begin{document}

\title{\vspace*{1cm}
Soliton solutions to the fifth-order Korteweg--de Vries
equation \\
and their applications to surface and internal water waves\\} %

\author{K.R. Khusnutdinova$^{1)}$, Y.A. Stepanyants$^{2, 3)}$, and M.R. Tranter$^{1)}$ \blfootnote{Corresponding author:
Yury.Stepanyants@usq.edu.au}\\}

\affiliation{\vspace*{0.5cm} $^{1)}$ Department of Mathematical
Sciences,
Loughborough University, \\Loughborough LE11 3TU, United Kingdom; \\%
$^{2)}$ Faculty of Health, Engineering and Sciences, University of
Southern Queensland, \\Toowoomba, QLD, 4350, Australia and \\
$^{3)}$ Department of Applied Mathematics, Nizhny Novgorod State
Technical University n.a. R.E. Alekseev, \\Nizhny Novgorod,
603950, Russia.}


\begin{abstract}%
\vspace*{1cm} %
We study solitary wave solutions of the fifth-order Korteweg--de
Vries equation which contains, besides the traditional quadratic
nonlinearity and third-order dispersion, additional terms
including cubic nonlinearity and fifth-order linear dispersion, as
well as two nonlinear dispersive terms. An exact solitary wave
solution to this equation is derived and the dependence of its
amplitude, width and speed on the parameters of the governing
equation are studied. It is shown that the derived solution can
represent either an embedded or regular soliton depending on the
equation parameters. The nonlinear dispersive terms can
drastically influence the existence of solitary waves, their
nature (regular or embedded), profile, polarity, and stability
with respect to small perturbations. We show, in particular, that
in some cases embedded solitons can be stable even with respect to
interactions with regular solitons. The results obtained are
applicable to surface and internal waves in fluids, as well as to
waves in other media (plasma, solid waveguides, elastic media with
microstructure, etc.).
\end{abstract}


\maketitle

\vfill

\clearpage

\section{Introduction}
\label{Intro} %

The Korteweg--de Vries (KdV) equation
\begin{equation}
\pdiff{u}{t} + c \pdiff{u}{x} + \alpha u \pdiff{u}{x} + \beta \pdiffn{3}{u}{x} = 0,
\label{KdV}
\end{equation}
is the well-known model for the description of weakly-nonlinear
long waves in media with small dispersion (see, for instance,
\cite{Karpman75,Whitham74,Lamb80,Ablowitz81,Dodd82,Newell85}). It
is widely used in the theory of long internal waves where it
describes astonishingly well the main properties of nonlinear
waves, even when their amplitudes are not small (see, for
instance, the reviews \cite{Ostrovsky89,Ostrovsky05,Apel07}). This
is the simplest model that combines the typical effects of
nonlinearity and dispersion, and provides stationary solutions
describing both periodic and solitary waves. The KdV equation is
completely integrable and possesses many remarkable properties,
which can be found in the references cited above.

\begin{figure}[ht!]
\includegraphics[width=0.9\textwidth]{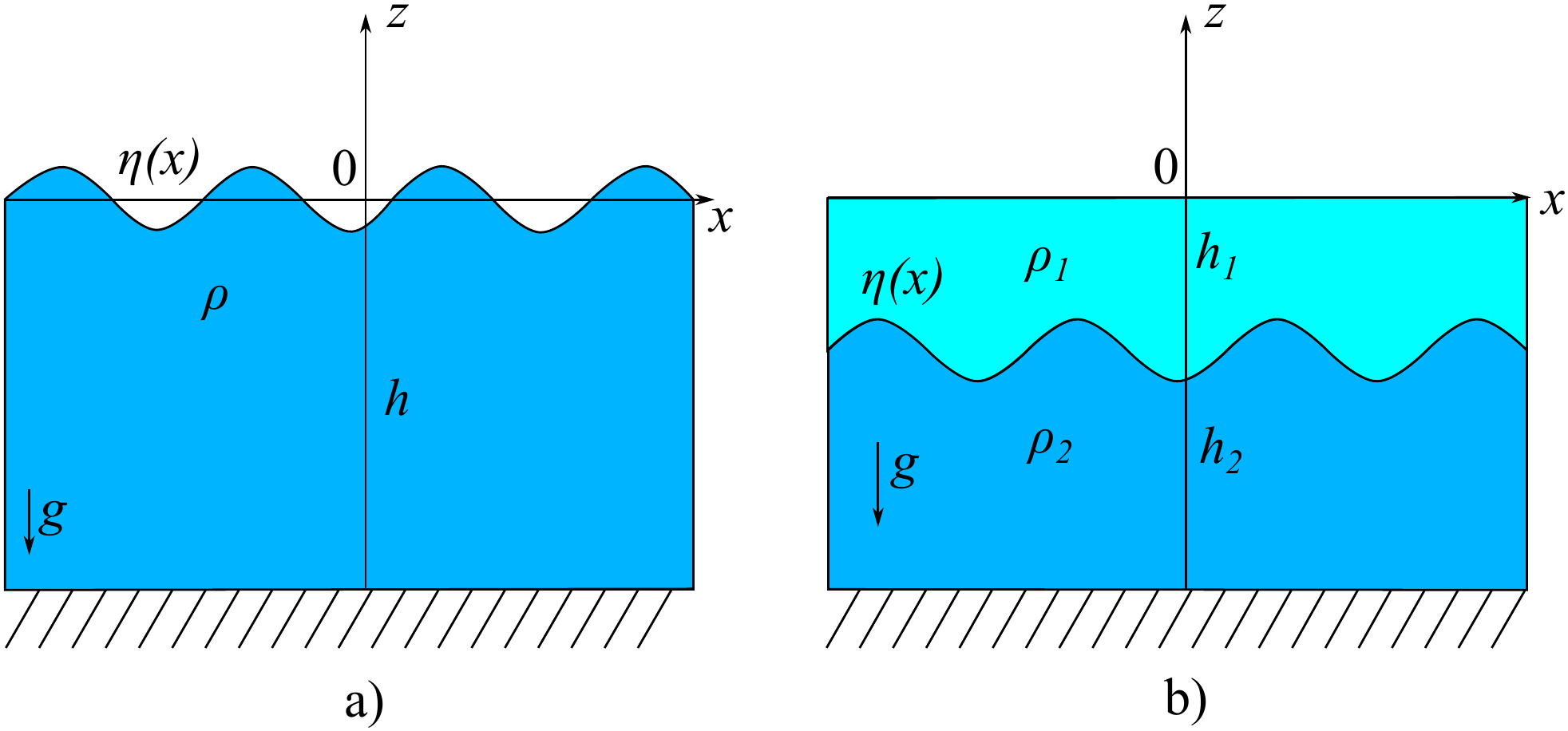}
\caption{Sketch of two configurations for surface water waves (a)
and internal waves in two-layer fluids (b). (Notice that the
horizontal and vertical scales are different.)} %
\label{fig:Sketch}
\end{figure}

At the same time, the KdV model cannot provide a detailed
description of many important features of nonlinear waves observed
in laboratory experiments, such as the non-monotonic dependence of
solitary wave speed on amplitude, or the table-top shape of
large-amplitude solitary waves \cite{Michallet&Barthelemy98}. To
capture such properties, the first natural step is a
straightforward extension of the KdV model by retaining the
next-order nonlinear and dispersive terms in the asymptotic
expansion of the solutions to primitive equations, for example the
Euler equations with boundary conditions appropriate for
oceanographic applications in the case of the ocean gravity waves.
A rather general form of the extended KdV equation has been
derived by many authors (in application to surface and internal
waves as shown in Figure \ref{fig:Sketch} see, e.g.,
\cite{Benney66, Lee74, Koop81, Olver84, MarchantSmyth90, Lamb96,
Grimshaw02, Giniyatullin14, KarRozRut14, Karczewska14}):
\begin{equation}
\pdiff{u}{t} + \alpha u \pdiff{u}{x} + \beta \pdiffn{3}{u}{x} + \varepsilon \lb \alpha_1 u^2 \pdiff{u}{x} + \gamma_1 u \pdiffn{3}{u}{x} + \gamma_2 \pdiff{u}{x} \pdiffn{2}{u}{x} + \beta_1 \pdiffn{5}{u}{x} \rb = 0.
\label{KdV2K}
\end{equation}
This equation, written in the coordinate frame moving with the
speed $c$, combines the quadratic ($\sim \alpha$) and cubic ($\sim
\alpha_1$) nonlinear terms, linear dispersion of the
3$^{\mathrm{rd}}$ ($\sim \beta$) and 5$^{\mathrm{th}}$ ($\sim
\beta_1$) orders, and also higher-order nonlinear dispersion terms
with coefficients $\gamma_1$ and $\gamma_2$; the parameter
$\varepsilon \ll 1$ is presumed to be small.

Particular cases of the fifth-order KdV equation, where some
coefficients are zero, were also derived for plasma waves
\cite{Kakutani69}, electromagnetic waves in discrete transmission
lines \cite{Gorshkov76}, gravity-capillary water waves
\cite{Abramyan85, Hunter88}, waves in a floating ice sheet (see
Ref. \cite{Guyenne14} and references therein).

In general equation \eqref{KdV2K} is not integrable, but for
particular choices of coefficients it reduces to one of a set of
equations that are completely integrable. These are the Gardner
equation \cite{Slyunyaev99, Slyunyaev01} (when $\beta_1 = \gamma_1
= \gamma_2 = 0$) or its particular case the standard KdV/mKdV
equation (when either $\alpha_1 = 0$ or $\alpha = 0$), as well as
the Sawada--Kotera and Kaup--Kupershmidt equations (when $\alpha =
\beta = 0$) \cite{Dodd82, Newell85}. A comprehensive discussion of
equation \eqref{KdV2K} and its properties can be found in
\cite{Kichenassamy92}.

The coefficients of equation \eqref{KdV2K} for surface gravity
waves are
\begin{equation}
c = \sqrt{gh}, \quad \alpha = \frac{3c}{2h},  \quad \alpha_1 = -\frac{3c}{8h^2}, \quad \beta = \frac{ch^2}{6}, \beta_1 = \frac{19ch^4}{360}, \quad \gamma_1 = \frac{5ch}{12}, \quad \gamma_2 = \frac{23ch}{24}.
\label{A9}
\end{equation}
For the gravity-capillary surface waves, as well as for internal
waves in a two-layer fluid, the coefficients are presented in
Appendix \ref{AppendA}. All notations are shown in Figure 1.

Unlike the KdV equation, the higher-order model \eqref{KdV2K} is
not a Hamiltonian equation and does not preserve the energy, in
general. However, in the particular cases when it reduces to
completely integrable models, it clearly becomes Hamiltonian.
Besides those cases, there is one more particular case of
$\alpha_1 = 0$ and $\gamma_2 = 2\gamma_1$ when equation
\eqref{KdV2K} becomes Hamiltonian, but nonintegrable
\cite{Champneys97, Yang01}. In the meantime, with the help of a
{\it near-identity transformation}, it can be mapped {\it
approximately} into one of a number of Hamiltonian equations
\cite{Kodama85, Fokas95, FokasLiu96, LiZhiSib97}. In particular
the asymptotic near-identity transformation
\begin{equation}
\tilde u = u + \varepsilon (a u^2 + b u_{xx})
\label{NIT}%
\end{equation}
maps equation \eqref{KdV2K} into itself up to terms of $o(\varepsilon)$, but with the new coefficients
\begin{equation*}
\tilde{\alpha_1} = \alpha_1 - a \alpha, \quad  \tilde{\beta_1} = \beta_1, \quad \tilde{\gamma_1} = \gamma_1, \tilde{\gamma_2} = \gamma_2 - 6 a \beta + 2 b \alpha,
\end{equation*}
where $a$ and $b$ are arbitrary parameters. If we choose the parameters $a = \lb \gamma_2 - 2 \gamma_1 \rb/6 \beta$ and $b = 0$, then equation \eqref{KdV2K} can be presented in the Hamiltonian form
\begin{equation}
u_t = \pdiff{ }{x} \lb \frac{\delta H}{\delta u} \rb,
\label{HamiltForm}
\end{equation}
where the Hamiltonian is $H = \int \mathcal{H} \dd{x}$ with the
density
\begin{equation}
\mathcal{H} = -\frac{1}{6} \alpha u^3 + \frac{1}{2} \beta u_{x}^{2}  - \varepsilon \lb \frac{1}{12} \tilde{\alpha_1} u^4 + \frac{1}{2} \tilde{\beta_1} u_{xx}^{2} - \frac{1}{2} \tilde{\gamma_1} u u_{x}^{2} \rb.
\label{Hamiltonian}
\end{equation}
The Hamiltonian form provides conservation of the ``mass'' $I_1 =
\int u \dd{x}$, ``wave energy'' $I_2 = \int \lb u^{2}/2 \rb
\dd{x}$, and Hamiltonian $I_3 = \int \mathcal{H} \dd{x}$. These
conserved quantities are very useful in the development of
asymptotic methods and perturbation techniques, as well as helping
to control the accuracy of numerical schemes. Notice however, as
has been shown in \cite{Olver84}, the formal Hamiltonians which
follow from the approximate evolution equations are not usually
the genuine Hamiltonians that can be derived from the primitive
equations for small-amplitude wave perturbations and which agree
with the physical energy conservation. Even in the classical KdV
equation \eqref{KdV} the Hamiltonian does not represent the
genuine wave energy. To this end the ``correct'' KdV equation with
the genuine Hamiltonian was derived in \cite{Olver84}; the
corresponding equation is a particular case of equation
\eqref{KdV2K} with $\beta_1 = 0$.

If the leading-order evolution equation is integrable, then the
underlying physical system is said to be asymptotically integrable
up to $O(\varepsilon)$. It turns out that for the higher-order KdV
equation \eqref{KdV2K} with special choices of coefficients, it is
possible to extend the asymptotic integrability, even up to
$O(\varepsilon^2)$ \cite{Kodama85, Fokas95, FokasLiu96}. Indeed,
the generic KdV equation is asymptotically reducible to the
integrable equation by the nonlocal near-identity transformation
\begin{equation}
\tilde{u} = u + \varepsilon \lb a_1 u^{2} + b_1 u_{xx} + c_1 u_{x} \int_{x_0}^{x} u \dd{\tilde{x}} + d_1 x u_{t} \rb,
\label{NIT1}
\end{equation}
where $a_1$, $b_1$, $c_1$, $d_1$ are arbitrary constants. This transformation can reduce \eqref{KdV2K} either to the next member of  the KdV hierarchy, or even to the classical KdV equation, with accuracy up to $O(\varepsilon^2)$. In particular, to transfer equation \eqref{KdV2K} to the KdV equation one should choose coefficients in \eqref{NIT1} of the form
\begin{align*}
&a_1 = \frac{-18 \beta^2 \alpha_1 + 2 \alpha^2 \beta_1 + 3 \alpha \beta \gamma_1}{9 \alpha \beta^2}, &&b_1 = \frac{-6 \beta^2 \alpha_1 - \alpha^2 \beta_1 + \alpha \beta \gamma_2}{2 \alpha^2 \beta}, \\
&c_1 = \frac{4 \alpha \beta_1 - 2 \beta \gamma_1}{9 \beta^2}, &&d_1 = - \frac{\beta_1}{3 \beta^2}.
\end{align*}
There are numerous other near-identity transformations; some of them are of special interest because they do not contain a secular term $\sim d_1$ as in the formula \eqref{NIT1}. In particular, the appropriately modified near-identity transformations reducing the higher-order KdV equation to the KdV equation have been successfully used to obtain particular solutions for the higher-order KdV equation from the known solutions of the KdV equation (e.g., the two-soliton solution extending the relevant KdV solution \cite{Kraenkel98, MarchantSmyth96, Marchant99} and the undular bore solution \cite{MarchantSmyth2006}).

In some particular cases when equation \eqref{KdV2K} is non-integrable it possesses, nevertheless, stationary solitary-type solutions, which can be constructed either numerically or sometimes even analytically. One of the best known cases is the Kawahara equation which follows from equation \eqref{KdV2K} when $\alpha_1 = \gamma_1 = \gamma_2 = 0$ \cite{Kawahara72}. This equation contains a rich family of solitary solutions including solitons with monotonic \cite{Yamamoto81} and oscillatory tails \cite{Kawahara72, Gorshkov76, Gorshkov79}.

Recently one more particular case of equation \eqref{KdV2K} was considered, the so--called Gardner--Kawahara equation \cite{Giniyatullin14, Kurkina15}, when only the nonlinear dispersive terms are absent ($\gamma_1 = \gamma_2 = 0$). Such a situation may occur, for example, in a two-layer fluid with surface tension between the layers. Solitary solutions for that equation were constructed numerically \cite{Kurkina15} and it was shown that among them there are ``fat solitons'', similar to those seen in the Gardner equation, and solitons with oscillatory tails, such as in the Kawahara equation, as well as their combination -- fat solitons with oscillatory tails.

An analytical solitary wave solution of the non-integrable
equation \eqref{KdV2K} was found in \cite{Karczewska14} for a
special set of coefficients, when it is not reduced to the
Sawada--Kotera or Kaup--Kupershmidt equations. The obtained
solution does not contain any free parameters and represents an
example of a so-called ``embedded soliton''. Embedded solitons
co-exist with linear waves propagating with the same speed (they
are ``embedded'' into the continuous spectrum of linear waves,
whereas regular solitons can be called ``gap'' solitons, because
they exist when the soliton speed belongs to a gap in the phase
speed spectrum of a corresponding linearized system
\cite{YMKC01}). The term ``embedded soliton'' was introduced in
Ref. \cite{Yang99}, although such solitons were known since 1974
when the first analytical example was obtained in Ref.
\cite{Nishikawa74} and their stability was proven numerically in
Ref. \cite{Gorshkov83} (some information about that can be found
also in Ref. \cite{Petviashvili92}). As currently known
\cite{Yang10}, embedded solitons can be both stable and unstable
with respect to perturbations of small or even big amplitudes
depending on the particular governing equation or set of
equations.

Nevertheless, the general problem of existence of solitary wave
solutions of equation \eqref{KdV2K} with arbitrary coefficients
remains open so far, and this circumstance motivated our study.
Besides the pure academic interest the problem is also topical in
application to surface and internal waves of large amplitude in
the ocean. Equation \eqref{KdV2K} can be considered as the model
equation capable of describing typical features of large-amplitude
solitary waves with good accuracy (see, e.g.,
\cite{Michallet&Barthelemy98} where it was shown that even its
reduced version, the Gardner equation, provides solutions similar
to those which can be constructed within the fully nonlinear Euler
equations). We study the role of higher-order nonlinear dispersive
terms ($\sim \gamma_1, \gamma_2$) and their influence on the shape
and polarity of solitary wave solutions. By means of the
Petviashvili numerical method \cite{Petviashvili92, Pelinovsky04}
we construct stationary solutions of equation \eqref{KdV2K} and
categorise them in terms of dimensionless parameters. We then
numerically model non-stationary solutions using a pseudospectral
scheme similar to that used in \cite{Grimshaw08, Alias13, Alias14,
Khusnutdinova17}. We show that these solutions demonstrate
soliton-like properties in the course of their interaction with
only minor inelastic effect. We also found an exact analytical
solution to this equation in the general case without a
restriction on its coefficients. The solution represents either
the embedded or a regular (gap) soliton, depending on parameters.

\section{Dimensionless form of the fifth-order KdV equation and its general properties}%
\label{Sect1}%
To minimise the number of parameters in the problem let us present equation \eqref{KdV2K} in dimensionless form using the change of variables
\begin{equation}
\tau = \frac{s\alpha^3 t}{\varepsilon \alpha_1} \sqrt{\frac{s}{\varepsilon \alpha_1 \beta}}, \quad
\xi = \alpha x \sqrt{\frac{s}{\varepsilon \alpha_1 \beta}}, \quad \upsilon = \frac{\varepsilon \alpha_1}{\alpha} s u,
\label{KdVCoV}
\end{equation}
where $s = \mathrm{sign}\lb \alpha_1 \beta \rb$ i.e. $s = 1$ if
$\alpha_1 \beta > 0$, and $s = -1$ if $\alpha_1 \beta < 0$. After
that the main equation \eqref{KdV2K} can be presented in the
divergent form:
\begin{equation}
\pdiff{\upsilon}{\tau} + \pdiff{ }{\xi} \lsq \frac{\upsilon^2}{2} + s \frac{\upsilon^3}{3} + \pdiffn{2}{\upsilon}{\xi} + B \pdiffn{4}{\upsilon}{\xi} + \frac{G_1}{2} \pdiffn{2}{\upsilon^2}{\xi} + \frac{G_2 -
3G_1}{2} \lb \pdiff{\upsilon}{\xi} \rb^2 \rsq = 0,
\label{DivForm}
\end{equation}
where
\begin{equation}
B = \alpha^2 \frac{s}{\alpha_1 \beta} \frac{\beta_1}{\beta}, \quad G_1 = \frac{s}{\alpha_1 \beta} \alpha\gamma_1, \quad G_2 = \frac{s}{\alpha_1 \beta} \alpha \gamma_2,
\label{DLSPar}
\end{equation}
(notice that $s/(\alpha_1 \beta) > 0$). The divergent form
immediately provides the ``mass conservation'' integral $I_1$ as
defined before. Multiplying this equation by $\upsilon$ and
integrating either over the entire axis $\xi$ for solitary waves
or over a period for periodic waves, we derive the ``energy
balance equation''
\begin{equation}
\pdiff{I_2}{\tau} \equiv \pdiff{ }{\tau} \int \frac{\upsilon^2}{2} \dd{\xi} = \lb G_1 - \frac{G_2}{2} \rb \int \lb \pdiff{\upsilon}{\xi} \rb^3 \dd{\xi}.
\label{EnBal}
\end{equation}
This relationship denotes a distinguishing feature of the model
\eqref{DivForm} which describes, in general, either time decay or
growth of ``wave energy'' $I_2$ due to the presence of the
nonlinear dispersive terms. Such conditionally defined ``wave
energy'' is conserved either in the trivial case when $G_1 = G_2 =
0$ or in the special case when $G_2 = 2G_1$ \cite{Champneys97}.
For stationary waves described by even functions $\upsilon(\xi)$,
the right-hand side of equation \eqref{EnBal} is zero, and their
``energy'' is also conserved for any values of $G_1$ and $G_2$.
(Note that in the case of surface gravity waves, $G_1 - G_2/2 =
-3/2$.)

\begin{figure}[!htbp]%
\centering
\includegraphics[width=0.5\textwidth]{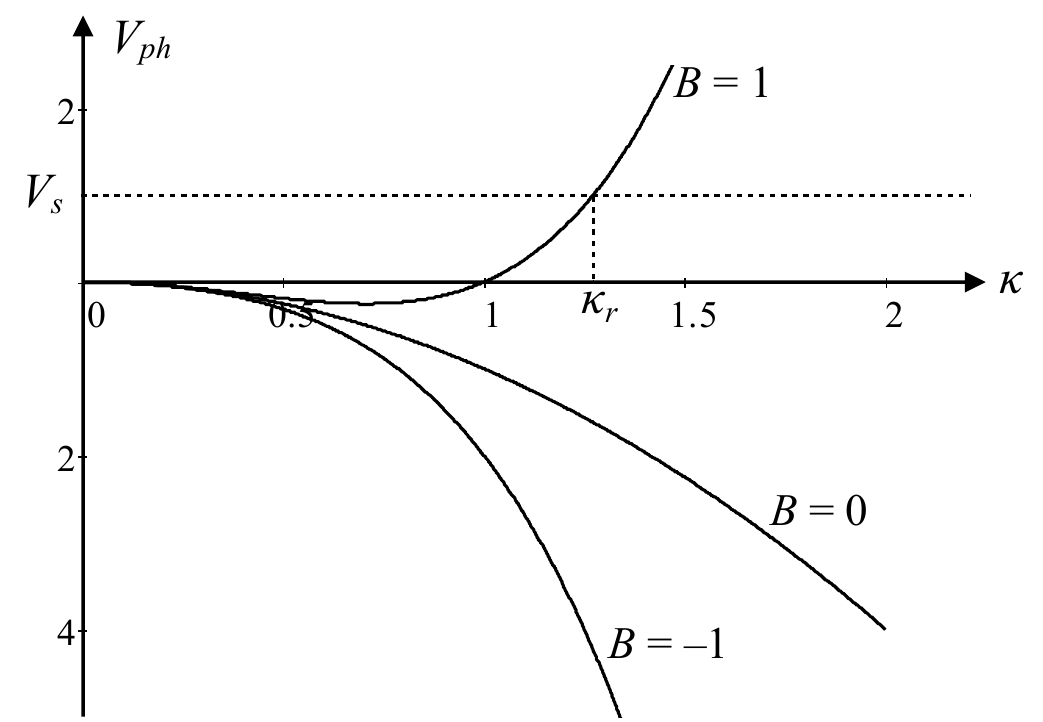} %
\caption{Plots of phase speed as per equation \eqref{DispRel} for zero, positive and negative values of parameter $B = 0$. The dashed horizontal line illustrates speed $V_s$ of a solitary wave, which can generate a small amplitude oscillatory wave at the resonant wavenumber $\kappa_r$.}
\label{fig:PhaseSpeeds}
\end{figure}

For waves of infinitesimal amplitude, $\upsilon \rightarrow 0$,
equation \eqref{DivForm} can be linearised. Looking for a solution
in the form $\upsilon \sim \exp{i \lb \tilde{\omega} \tau - \kappa
\xi \rb}$, we obtain the dispersion relation $\tilde{\omega} \lb
\kappa \rb$ and phase speed $V_{ph}(\kappa)$, in the coordinate
frame moving with speed $c$, of the form
\begin{equation}
\tilde{\omega} = -\kappa^3 + B \kappa^5, \quad V_{ph}(\kappa) \equiv \frac{\tilde{\omega}}{\kappa} = -\kappa^2 + B \kappa^4.
\label{DispRel}
\end{equation}
The plot of the phase speed is shown in Figure
\ref{fig:PhaseSpeeds} for three typical values of the parameter
$B$. Note that for surface water waves $B = 57/5 > 0$, therefore
qualitatively the dispersion curve is similar to curve $B = 1$.
The same is true for oceanic internal waves, as follows from the
expressions for the coefficients of \eqref{KdV2K} (see Appendix
\ref{AppendA}). In some cases for internal waves in laboratory
tanks the coefficient $B$ can be zero or negative.

In the case of $B \leq 0$ the phase speed is a monotonic function
of $\kappa$, whereas for $B > 0$ it has a minimum, $V_{min} =
-1/(4B)$, at the point $\kappa_c = \lb 2 B \rb^{-1/2}$ (see Figure
\ref{fig:PhaseSpeeds}). The concept of phase speed is very useful
in understanding the process of interaction of a moving source
(e.g. a solitary wave) with linear waves. In particular, if the
speed of a source $V_s$ is such that there is no resonance with
any linear wave i.e. there is no intersection of the dashed line
in Figure \ref{fig:PhaseSpeeds} with the dispersion curve (e.g.
when $V_s < V_{\mathrm{min}}$), then the source does not lose
energy for the wave excitation. Otherwise, in the case of a
resonance (see the intersection of the dashed line with the curve
for $B = 1$) the source, in general, can experience energy losses
for the generation of a linear wave and, as a result, it gradually
decelerates. Without external compensation of energy losses, such
a source usually cannot move steadily. However, there are several
examples of embedded solitons which can steadily propagate with
the same speed as a linear wave, but not exciting it effectively.
The physics of such a phenomenon has not been well understood yet;
it requires further study which is beyond the scope of this paper.
Therefore, the no-resonance condition $V_{s} \neq V_{ph}$ can
provide only an indication of when a solitary wave can most likely
be expected.

\section{Stationary solutions of the fifth-order KdV equation}%
\label{Sect2}%
Consider now stationary solutions to equation \eqref{DivForm} in the form of travelling waves depending only on one variable, $\zeta = \xi - V \tau$, where $V$ is the wave speed. In this case equation \eqref{DivForm} can be reduced to an ODE and integrated once (the constant of integration is set to zero for solitary waves):
\begin{equation}
B \diffn{4}{\upsilon}{\zeta} + \diffn{2}{\upsilon}{\zeta} - V \upsilon + \frac{\upsilon^2}{2} + s \frac{\upsilon^3}{3} + \frac{G_1}{2} \diffn{2}{\upsilon^2}{\zeta} + \frac{G_2 - 3G_1}{2} \lb
\diff{\upsilon}{\zeta} \rb^2 = 0.
\label{SolODE}
\end{equation}
There are five independent parameters in this equation: $B$,
$G_1$, $G_2$, $s$, and $V$, which determine the structure of a
solitary wave. This equation actually splits into two independent
equations with different properties; one equation with negative
cubic term ($s = -1$), and another one with positive cubic term
($s = 1$). We will derive a particular soliton solution to this
equation for both $s=1$ and $s=-1$. The case of $s = 0$ with a
specific link between the parameters, $G_2 = 2G_1$, has been
studied in \cite{Champneys97}.

As a first step, let us consider solitary wave asymptotics at
plus/minus infinity. Assuming that soliton solutions decay at
infinity, let us linearise equation \eqref{SolODE} (simply omit
all nonlinear terms) and seek a solution of the remaining linear
equation in the form $\upsilon \sim \exp{\lb \mu \zeta \rb}$.
Substituting this trial solution into the linearised equation
\eqref{SolODE}, we obtain an algebraic equation for $\mu$ of the
form (cf. \cite{Champneys97}):
\begin{equation}
B \mu^4 + \mu^2 - V = 0.
\label{AlgEqForMu}
\end{equation}
The roots of this bi-quadratic equation are
\begin{equation}
\mu_{1,2} = \pm \sqrt{\frac{-1 + \sqrt{1 + 4BV}}{2B}}, \quad \mu_{3,4} = \pm \sqrt{\frac{-1 - \sqrt{1 + 4BV}}{2B}}.
\label{RootsBQ-Eq} %
\end{equation}
Let us analyse the roots in detail (an alternative analysis of the
roots in the plane of different parameters can be found in
\cite{Champneys97}). Firstly we consider the case when the
parameter $B$ is negative. For negative $V$ we have $4BV
> 0$ and $\sqrt{1 + 4BV} > 1$, therefore the roots $\mu_{1,2}$ are
purely imaginary, and the roots $\mu_{3,4}$ are real. Solutions
corresponding to purely imaginary roots are not decaying and
cannot represent solitary waves with zero amplitude at infinity.
If $V = 0$, then $\mu_{1,2} = 0$ and the corresponding solutions
do not decay at infinity.

If $0 < V < -1/(4B)$, then we have $\sqrt{1 + 4BV} > 0$, and all four roots $\mu_{1,2,3,4}$ are real. In this case soliton solutions are possible, with exponentially decaying asymptotics at infinity. Finally, if $V > -1/(4B)$, then $\sqrt{1 + 4BV}$ is complex; all roots are complex-conjugate in pairs $\mu_{1,2} = \pm (p_1 + iq_1)$, $\mu_{3,4} = \pm (p_2 + iq_2)$. Due to the presence of the real parts of the roots, $p_{1,2}$, there can be soliton solutions with oscillatory asymptotics. The decay rate of a solitary wave in the far field is determined by the root with the smallest value of $|p_{1,2}|$.

Assume now that $B$ is positive. It follows from a similar analysis of roots as was done for negative $B$ that, for $V < V_{min} \equiv -1/(4B)$, solitary waves with oscillatory asymptotics are possible. For $V_{min} < V < 0$, the roots are purely imaginary, therefore no solitons with zero asymptotics can exist in this case. For $V > 0$ the roots
$\mu_{1,2}$ are purely real, then embedded solitons with exponential asymptotics are possible.

\begin{figure}[!htbp]%
\includegraphics[width=15cm]{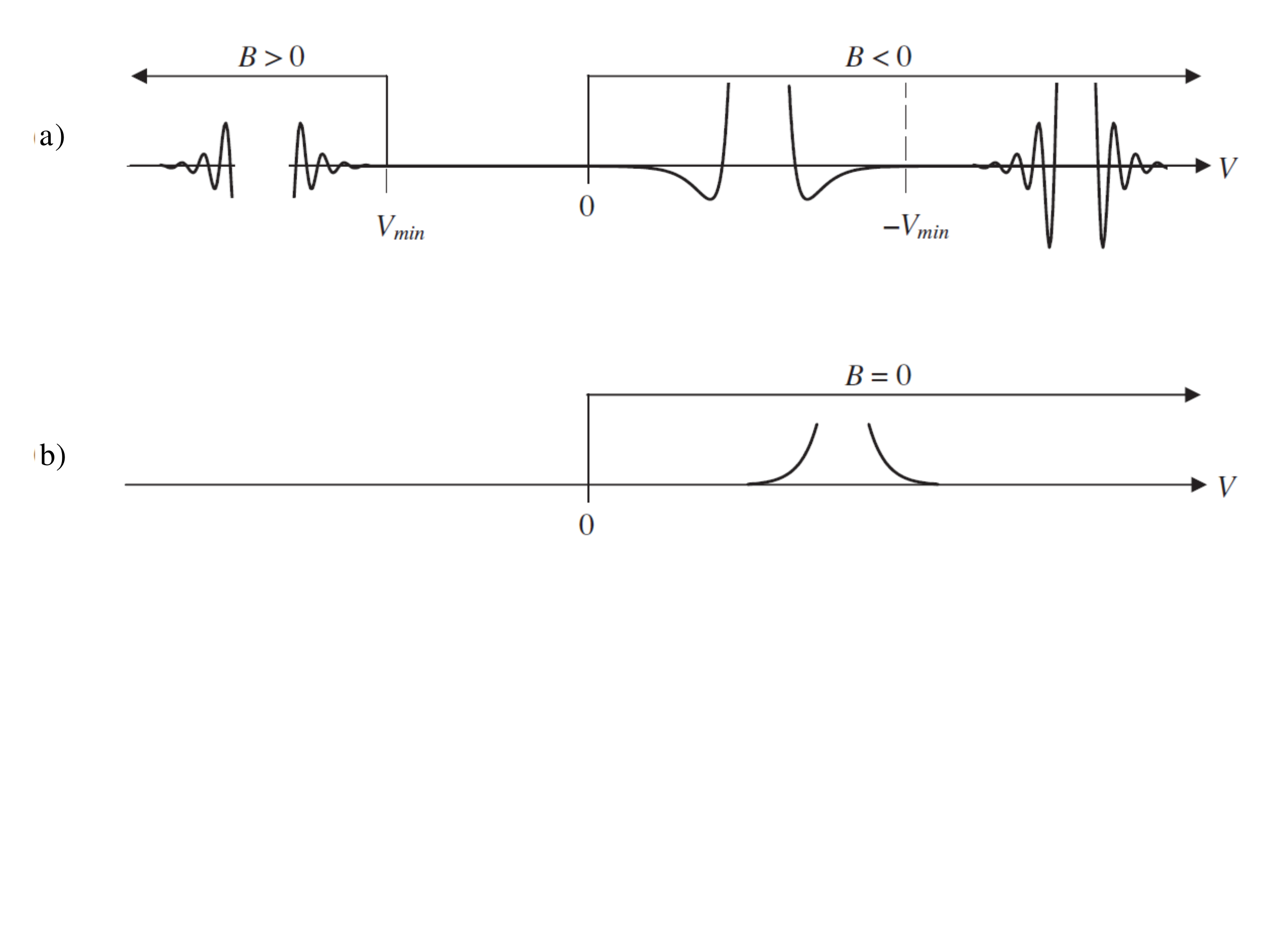} %
\vspace*{-4.5cm}%
\caption{Asymptotics of possible soliton solutions of equation
\eqref{SolODE} with different signs of parameter $B$ (panel a).
Panel (b) shows the asymptotics of regular solitons in the case of
$B = 0$ or embedded solitons in the case
of $B > 0$.} %
\label{fig:RootSchem}
\end{figure}

In the particular case of $B = 0$, equation \eqref{AlgEqForMu} has two real roots corresponding to soliton solutions, provided that $V > 0$. These findings can be summarised with the help of a schematic diagram, as shown in Figure \ref{fig:RootSchem}.

It should be noted that the analysis of roots only predicts \textit{possible} asymptotics of solitons provided that they exist, but it does not guarantee their existence. In particular, if $B = 0$, then soliton solutions with monotonically decaying exponential asymptotics could exist for any $V > 0$ (see case (b) in the diagram), but in fact they exist only for $0 < V < 1/6$ (see, e.g., \cite{Kurkina15}).

\subsection{Solitary wave solution of the fifth-order KdV equation}
\label{Subsec2.1}%
Let us consider a trial solution to \eqref{SolODE} in the form of a sech$^2$ solitary wave (in a similar way to what was done in \cite{Kichenassamy92}), taking the form
\begin{equation}
\upsilon \lb \zeta \rb = A \sechn{2}{K \zeta},
\label{TrialSol}
\end{equation}
where $A$ is the soliton amplitude and $K$ is the parameter
determining its half-width $\Delta = 1/K$ (similar solution was
constructed in \cite{Champneys97} for the particular case of
$s = 0$ and $G_2 = 2G_1$). By substitution of this solution into
equation \eqref{SolODE} we obtain
\begin{equation}
C_2 \sechn{2}{K \zeta} + C_4 \sechn{4}{K \zeta} + C_6 \sechn{6}{K \zeta} = 0,
\label{SubTrial}
\end{equation}
where
\begin{align}
C_2 &= 16BK^4 + 4K^2 - V, \label{C2Coef} \\
C_4 &= -120BK^4 + [2(G_1 + G_2)A - 6]K^2 + \frac{A}{2}, \label{C4Coef} \\
C_6 &= 120BK^4 + s \frac{A^2}{3} - 2(G_2 + 2G_1)AK^2 %
\label{C6Coef}
\end{align}
(cf. \cite{Kichenassamy92} where a slightly different approach was used). Equating the coefficients $C_2$ and $C_4$ to zero, we obtain
\begin{align}
V &= 4K^2 \lb 1 + 4BK^2 \rb = \frac{4 \lb \Delta^2 + 4B \rb}{\Delta^4}, \label{VSol} \\
A &= 12K^2 \frac{1 + 20BK^2}{1 + 4 \lb G_1 + G_2 \rb K^2} = \frac{12}{\Delta^2} \frac{\Delta^2 + 20B}{\Delta^2 + 4 \lb G_1 + G_2
\rb}. \label{ASol}
\end{align}
Eliminating $A$ from equation \eqref{C6Coef} with the help of equation \eqref{ASol}, we obtain the quadratic equation for $R \equiv K^2$ of the form
\begin{equation*}
80B \lb G_1^2 + G_1 G_2 - 10Bs \rb R^2 + 4 \lb 2G_1^2 + 3 G_1 G_ 2 + G_2^2 - 5 BG_2 - 20Bs \rb R - 5B + 2G_1 + G_2 - 2s = 0.
\label{BiQuad}
\end{equation*}
This equation has two roots, in general, and the corresponding expression for the half-width of a solitary wave is determined by
\begin{equation}
\Delta_{1,2}^2 = \frac{40B \lb G_1 G_2 + G_1^2 - 10Bs \rb}{5B \lb G_2 + 4s \rb - 3G_1 G_2 - 2G_1^2 - G_2^2 \pm \lb 5B - G_1 - G_2 \rb \sqrt{\lb 2G_1 + G_2 \rb^2 - 40Bs}}.
\label{SolHW}
\end{equation}
Thus, we see that a solitary wave solution in the form of \eqref{TrialSol} does exist for a certain set of parameters $B$, $G_1$ and $G_2$. One of the obvious restrictions on the set of parameters is (we have two expressions as $s$ can be positive or negative)
\begin{equation}
\lb 2G_1 + G_2 \rb^2 - 40 B s \geq 0 \RA B \geq -\frac{\lb 2G_1 + G_2 \rb^2}{40} \eqtext{for $s = -1$, or} B \leq \frac{\lb 2G_1 + G_2 \rb^2}{40} \eqtext{for $s = 1$}.
\label{RestrB}
\end{equation}
Below we consider a few particular cases and analyse the corresponding soliton solutions.

\subsection{The Gardner--Kawahara equation ($G_1 = G_2 = 0$)}
\label{Subsubsec2.1.1}%
In this case the expressions for $\Delta$, $V$ and $A$ simplify considerably. The half-width is given by
\begin{equation}
\Delta^2 = \frac{-40Bs}{2s \pm \sqrt{-10Bs}}.
\label{G1G2Eq0}
\end{equation}
The soliton derived from \eqref{G1G2Eq0} can either be an embedded soliton or a regular soliton, dependent upon its speed as defined by \eqref{VSol} and the value of $B$. Due to the presence of $s$ in \eqref{G1G2Eq0} we consider two sub-cases: $s = -1$ and $s = 1$.

\subsubsection{The Gardner--Kawahara equation with $s = -1$}
In this case the only meaningful solution to \eqref{G1G2Eq0} is
\begin{equation}
\Delta^2 = \frac{40B}{-2 + \sqrt{10B}},
\label{G1G2Eq0sm1}
\end{equation}
with $B > 2/5$ for a real solution. Then from equation
\eqref{VSol} it follows that $V > 0$. With $B > 0$ the phase speed
dependence of linear waves on the wavenumber is shown in Figure
\ref{fig:PhaseSpeeds} by the upper line. Therefore, solution
\eqref{TrialSol} with $V > 0$ is the embedded soliton moving in
resonance with a linear wave (however gap solitons can co-exist
with the embedded soliton as will be shown below). Figure
\ref{fig:NegNonlin} shows the dependence of the soliton parameters
on $B$. The soliton amplitude monotonically increases with $B$,
whereas its width and speed non-monotonically depend on this
parameter. The minimum value $\Delta_{min} = \sqrt{32}$ is
attained at $B = 8/5$, and maximum speed $V_{max} = 5/32$ occurs
at $B = 128/45$. Soliton profiles for these two cases will be
shown below in Figure \ref{fig:GKsm1} in comparison with regular
solitons, numerically constructed for the same values of $B$. The
asymptotics of embedded solitons are in agreement with the
prediction from the analysis of roots of the characteristic
equation \eqref{AlgEqForMu} - see Figure \ref{fig:RootSchem}b.
\begin{figure}[!htbp]%
\includegraphics[width=12cm]{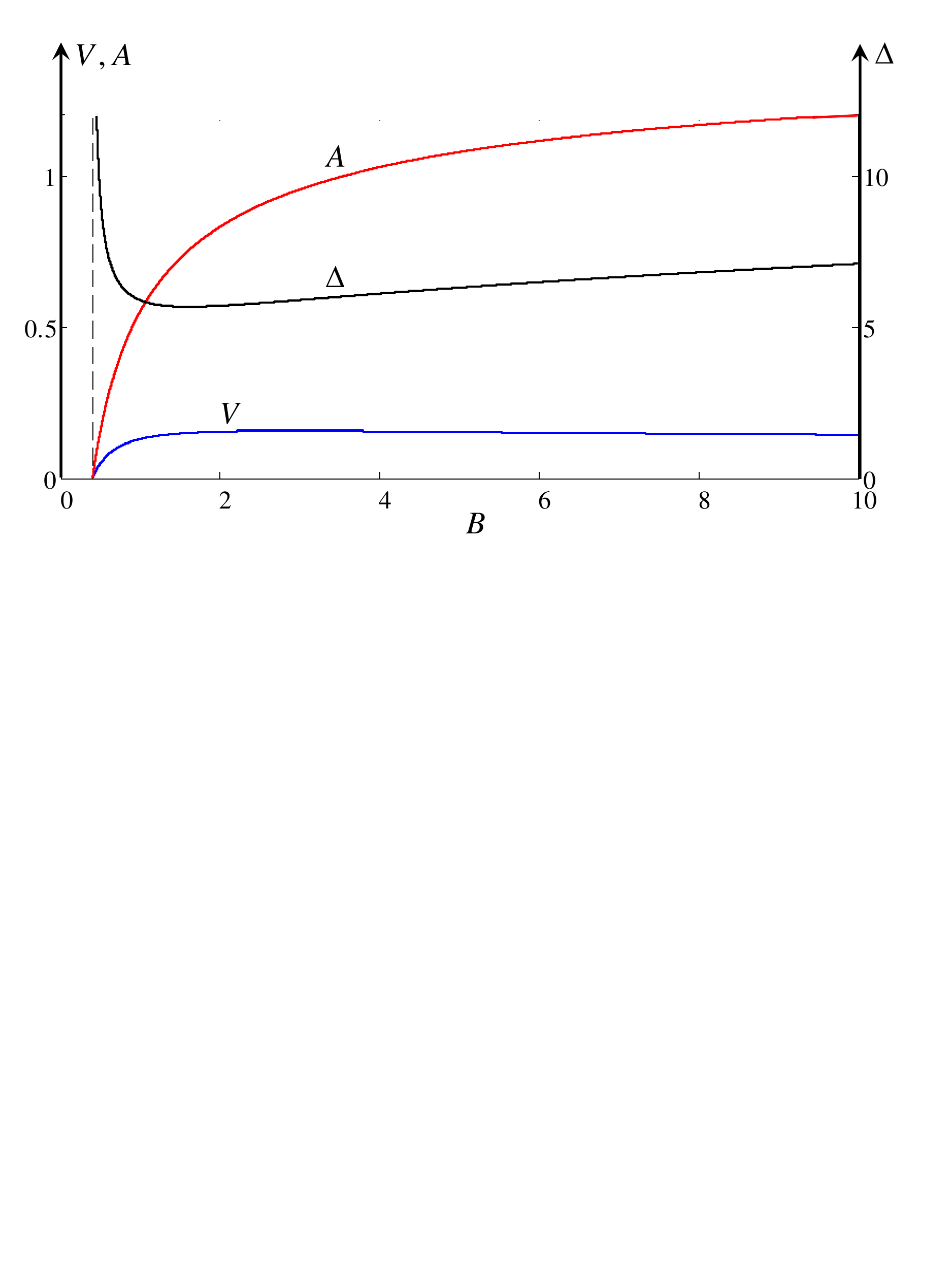} %
\vspace{-9.0cm}%
\caption{(color online) Dependence of embedded soliton parameters on $B$, for $G_1 = G_2 = 0$ with $s = -1$. Vertical dashed line shows the limiting value of the parameter $B_{lim} = 2/5$, below which the soliton does not exist.}
\label{fig:NegNonlin}
\end{figure}

\subsubsection{The Gardner--Kawahara equation with $s = 1$}
\begin{figure}[!htbp]%
\includegraphics[width=12cm]{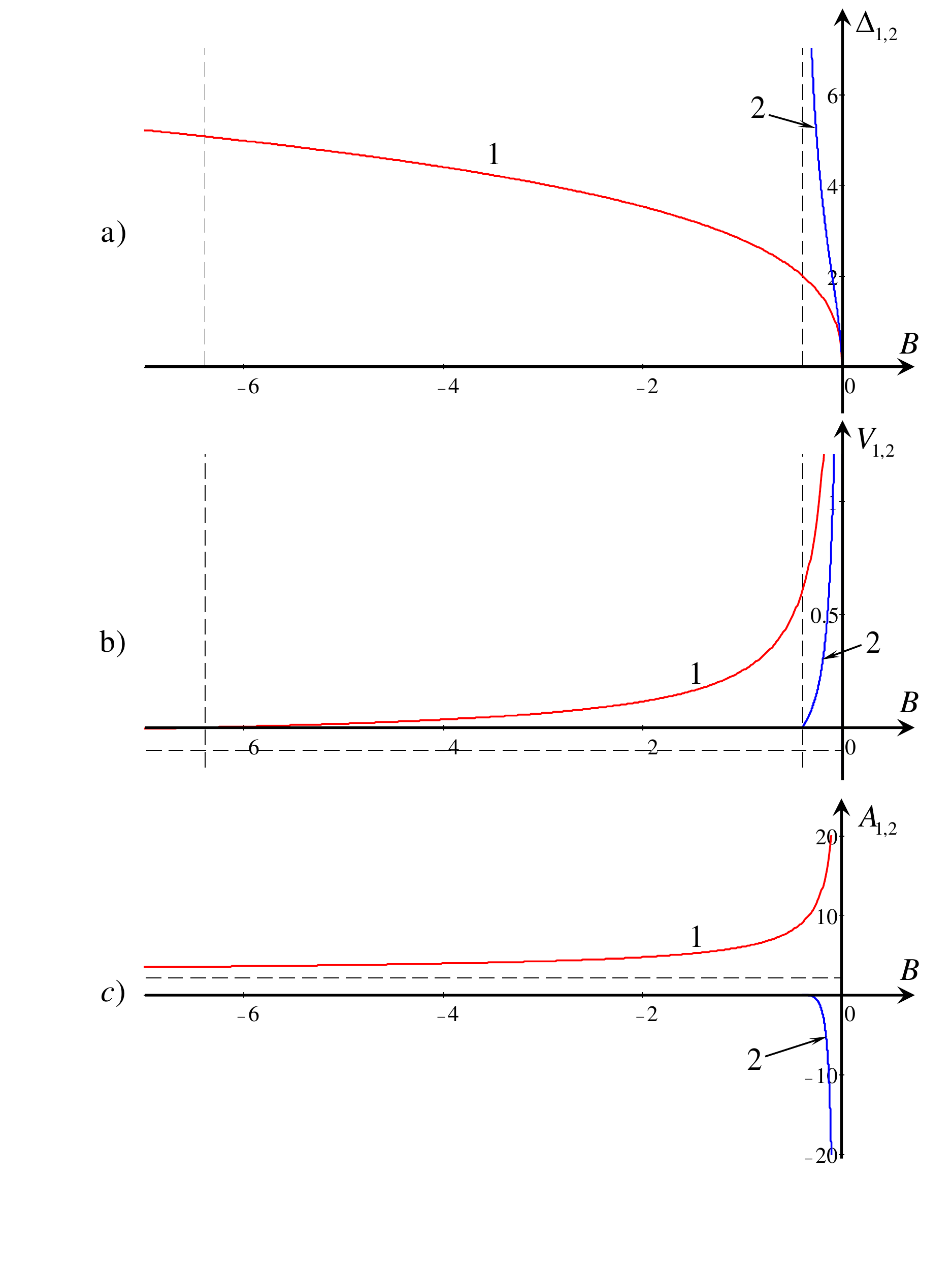}
\vspace{-1.0cm}%
\caption{(color online) Dependence of soliton parameters on $B$,
for $G_1 = G_2 = 0$ with $s = 1$. Frame a): widths of two
different solitons as per equation \eqref{G1G2Eq0sp1}; frame b):
speeds of two different solitons as per equation \eqref{VRestr};
frame c): amplitudes of two different solitons as per equations
\eqref{ASol1} and \eqref{ASol2}. Line 1 is for positive sign in
these expressions and line 2 for negative signs. Dashed vertical
lines correspond to the critical values of parameter $B$: $B =
-32/5$ (left line) and $B = -2/5$ (right line). The horizontal
dashed line in frame b) shows the limiting value of soliton speed
$V_{lim} = -1/10$, when $B \to -\infty$, and the horizontal dashed
line in frame c) shows the limiting value of soliton amplitude
$A_{lim} = 9/4$, when $B \to -\infty$.} %
\label{fig:PosNonlin}
\end{figure}

In this case the soliton half-width takes the form
\begin{equation}
\Delta^2 = \frac{-40B}{2 \pm \sqrt{-10B}}, \eqtext{where $B < 0$.}
\label{G1G2Eq0sp1}
\end{equation}
If the positive sign is taken in front of the square root, then the right-hand side of the expression is positive for all negative $B$, whereas if the negative sign is taken in front of the square root then the expression in the right-hand side is positive only under the restriction $B > -2/5$ (see Figure \ref{fig:PosNonlin} (a)).

Then, from Figure \ref{fig:PhaseSpeeds} it follows that for $B <
0$ the solution \eqref{TrialSol} represents a regular soliton if
$V > 0$, and an embedded soliton if $V < 0$. The soliton velocity
\eqref{VSol} with $\Delta^2$ given by equation \eqref{G1G2Eq0sp1}
is
\begin{equation}
V = \frac{4}{\Delta^4} \lb \Delta^2 + 4B \rb = -\frac{5B + 8 \pm 3 \sqrt{-10B}}{50B}.
\label{VRestr}%
\end{equation}
The analysis of this expression shows that, if the positive sign
is chosen in front of the square root, then solution
\eqref{TrialSol} represents a regular soliton with $V > 0$, if
$-32/5 < B < 0$, and an embedded soliton with $V < 0$, if $B <
-32/5$. If the sign in front of the square root is negative, then
solution \eqref{TrialSol} represents a regular soliton with $V >
0$ within the interval $-2/5 < B < 0$, whereas beyond this
interval soliton solutions do not exist (see Figure
\ref{fig:PosNonlin}b).

As follows from equation \eqref{ASol}, solitons with positive sign in front of the square root in equation \eqref{G1G2Eq0sp1} have positive polarity, and solitons with negative sign have negative polarity (see Figure \ref{fig:PosNonlin}c). In particular, when $G_1 = G_2 = 0$, we have
\begin{align}
A_1 &= -\frac{360B}{\Delta_1^4} = -\frac{9}{20B}\lb 2 - 5B + 2\sqrt{-10B} \rb, \label{ASol1} \\
A_2 &= -\frac{360}{\Delta_2^4} = -\frac{9}{20B^2}\lb 2 - 5B - 2\sqrt{-10B} \rb. \label{ASol2}
\end{align}
Vertical dashed lines in Figure \ref{fig:PosNonlin} at $B = -2/5$
show the limiting value of this parameter below which the negative
polarity solitons cannot exist. Other vertical dashed lines in
Figure \ref{fig:PosNonlin} at $B = -32/5$ show the boundary
between the regular solitons with $V > 0$ and $B
> -32/5$, and embedded solitons with $V < 0$ and $B < -32/5$.

The horizontal dashed line in frame b) shows the limiting value of the embedded soliton speed $V_{lim} = -1/10$, when $B \to -\infty$, and the horizontal dashed line in frame c) shows the limiting value of the embedded soliton amplitude $A_{lim} = 9/4$, when $B \to -\infty$ (the width of the embedded soliton slowly
increases when $B \to -\infty$ as $\Delta \sim 2\sqrt[4]{-10B}$).

\begin{figure}[h!]%
\includegraphics[width=16cm]{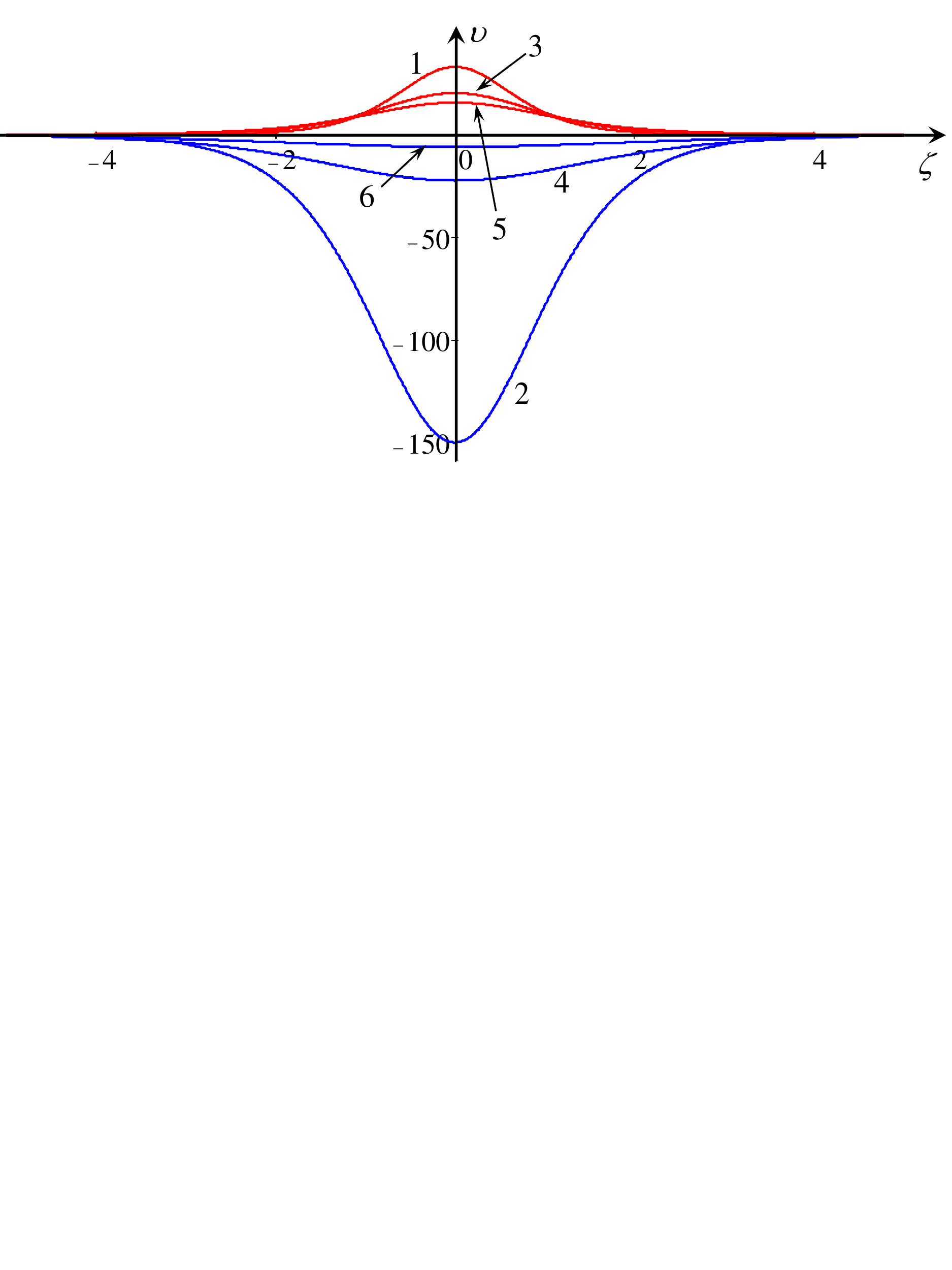} %
\vspace{-13.5cm}%
\caption{(color online) Profiles of gap solitons of positive and
negative polarities for three values of $B$ in the interval $-2/5
< B < 0$. Specifically $B = -0.05$ for lines 1 and 2, $B = -0.1$
for lines 3 and 4, and $B = -0.15$ for lines 5 and 6.} %
\label{fig:TwoSols}
\end{figure}

Thus, in the interval $-2/5 < B < 0$, two types of regular
solitons can coexist with different widths, speeds, amplitudes,
and polarities. When $B < -2/5$ only one soliton of positive
polarity can exist which smoothly transfers into the embedded
soliton when the parameter $B$ passes through the threshold value
$B = -32/5$ where the velocity vanishes. Figure \ref{fig:TwoSols}
illustrates the profiles of regular solitons of positive and
negative amplitudes for three values of $B$ in the indicated
interval.

\subsection{A particular case of surface gravity waves ($s = -1$)}
\label{Subsubsec2.1.2}%

Using the expressions for the coefficients of equation
\eqref{KdV2K} for gravity waves (see equation \eqref{A9}), we
obtain the dimensionless parameters $B = 57/5$, $G_1 = 10$, $G_2 =
23$ (notice that in this case $G_2 = 2.3G_1$, which is very close
to the case $G_2 = 2G_1$ when the energy is conserved -- see
equation \eqref{EnBal}). For this set of parameters there is only
one real root of equation \eqref{SolHW}
\begin{equation}
\Delta_{1} = \sqrt{\frac{8436}{\sqrt{2305} - 14}} \approx 15.75.
\label{D1}
\end{equation}
After that we find the amplitude and speed of a soliton in the form
\begin{equation}
A = \frac{3}{148}\left(51 - \sqrt{2305}\right) \approx 6.06 \cdot
10^{-2}, \quad V = \frac{157\sqrt{2305} - 89}{390165} \approx 1.91
\cdot 10^{-2}. \label{AV}
\end{equation}
We see again that $V > 0$ and $B > 0$, therefore the soliton is in
resonance with one of the linear waves as shown in Figure
\ref{fig:PhaseSpeeds} by the upper line. Hence, solution
\eqref{TrialSol} represents again the embedded soliton with
exponential asymptotics in agreement with the prediction following
from the analysis of roots of characteristic equation
\eqref{AlgEqForMu} - see Figure \ref{fig:RootSchem}b).

In dimensional variables the amplitude, width and speed of the surface embedded soliton are
\begin{equation*}
A_d = 4hA, \quad \Delta_d = h \Delta/6, \quad V_{d} = 6V \sqrt{gh},
\end{equation*}
where index $d$ pertains to the dimensional variables, and $h$ is
the fluid depth. Setting $h = 10$ m, we obtain $A_d \approx 2.42\
\text{m}$, $\Delta_d \approx 26.25 ~\text{m}$, $V_d \approx 1.14\
\text{m/s}$ (the total speed is $c + V_d \approx 9.9\ \text{m/s} +
1.14\ \text{m/s} \approx 11.04\ \text{m/s}$). Thus, according to
this solution, a soliton of small amplitude can exist on the
surface of the water. In the meantime, the classical KdV theory
predicts the existence of a soliton which has amplitude $A_e
\approx 5.37\ \text{m}$ and speed $V_e \approx 2.66\ \text{m/s}$
at the same width $\Delta_d \approx 26.25\ \text{m}$.

The linear wave of infinitesimal amplitude propagating with the
same speed as the embedded soliton ($V_{ph} = c + V_d$) has
dimensional wavelength
\begin{equation*}
\lambda_d = \pi \sqrt{\frac{2\beta}{V_d} \lb \sqrt{1 + \frac{4 \beta_1 V_d}{\beta^2}} - 1 \rb}.
\end{equation*}
Substituting the coefficients of equation \eqref{KdV2K} for pure gravity waves (see equation \eqref{A9} in Appendix \ref{AppendA}) and $h = 10$ m, we obtain $\lambda_d \approx 32.5 \mbox{m}$.

As mentioned above, the embedded solitons can be stable even with
respect to big perturbations. We will demonstrate in section
\ref{Sect4} that they can survive even after strong interactions
with regular solitons. The problem of general evolution of
embedded solitons under the action of small perturbations caused
by medium inhomogeneity or energy dissipation is still an open
problem and worth a further study. Some preliminary results can be
found in \cite{Karczewska14, Yang10}.

\subsection{A particular case when $G_2 = 3G_1$}
\label{Subsubsec2.1.3}%
One more special case of $G_2 = 3G_1$ is worth considering because, in this case, the basic equation \eqref{SolODE} simplifies. In this simplified form, we can use the numerical code described in Appendix \ref{AppendB}, based on the Petviashvili method and adapted for the solution of equation \eqref{SolODE} without the last term. This allows us to find the solutions numerically and compare the results obtained with the analytical solutions derived here. This case will also allow us to understand the role of nonlinear dispersion in the energy balance equation \eqref{EnBal}. According to that equation, the energy is conserved on even solutions, and therefore stationary solutions in the form of solitary waves may exist. However the question is, what happens when solitary waves interact? We consider this problem in Section \ref{Sect4} by direct numerical solution of the non-stationary equation \eqref{DivForm}.

\begin{figure}
    \centering
    \includegraphics[width = 0.8\textwidth]{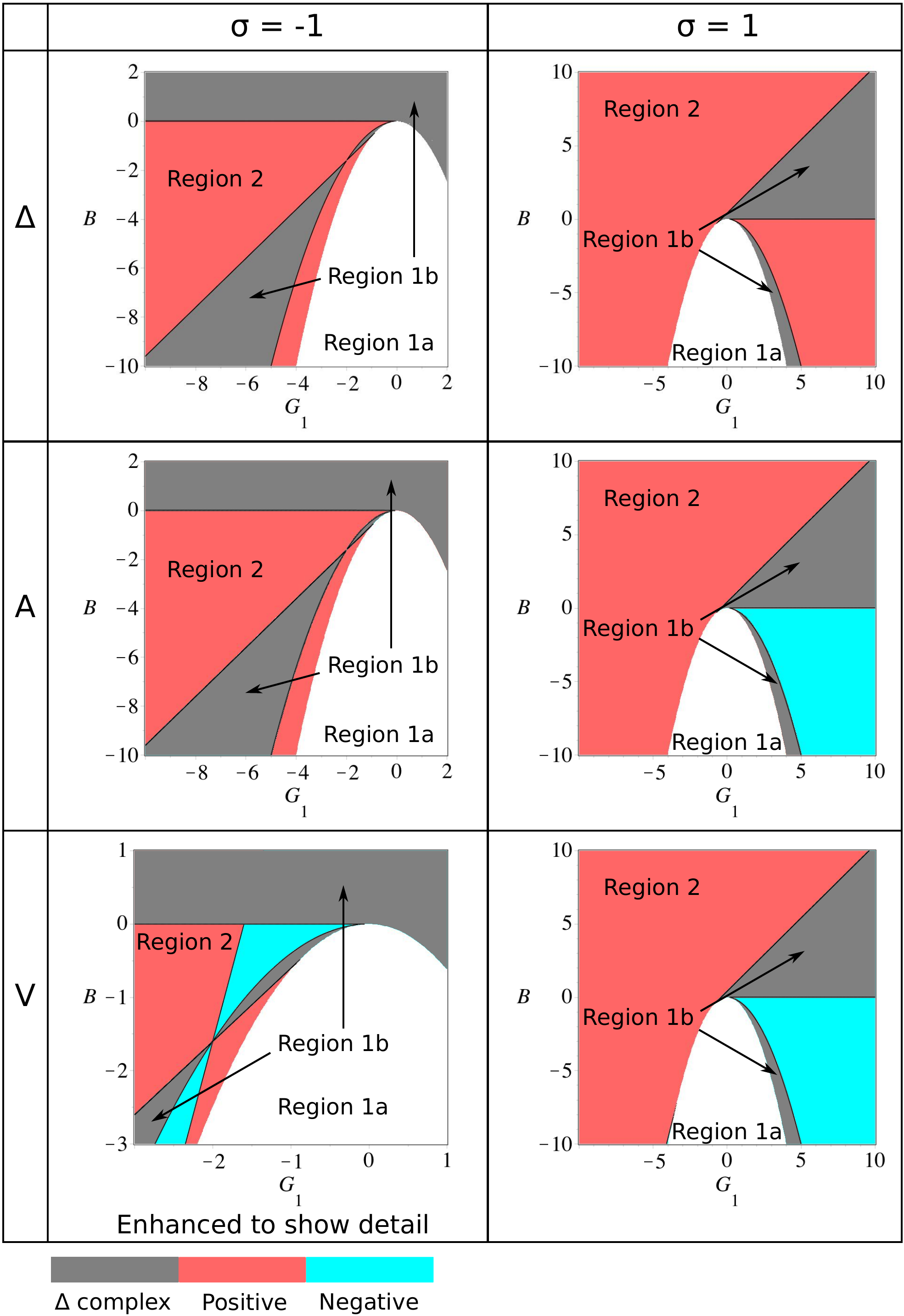}
    \caption{(color online) The parameter plane $(B, G_1)$ for the cases $\sigma = \pm 1$ in equation \eqref{SolHW13} where soliton solutions in the form of \eqref{TrialSol} can exist in region 2. Region 1a and region 1b denote areas where $\Delta$ is complex. Subsequent panels show subregions of region 2, where soliton speed or amplitude is positive (red) or negative (blue). The images for $\Delta$ use red shading for region 2, for reference. We use an enhanced scale in the lower left frame for $V$ to show detail.}
\label{fig:ESPlotsm1}
\end{figure}

\begin{figure}
    \centering
    \includegraphics[width = 0.8\textwidth]{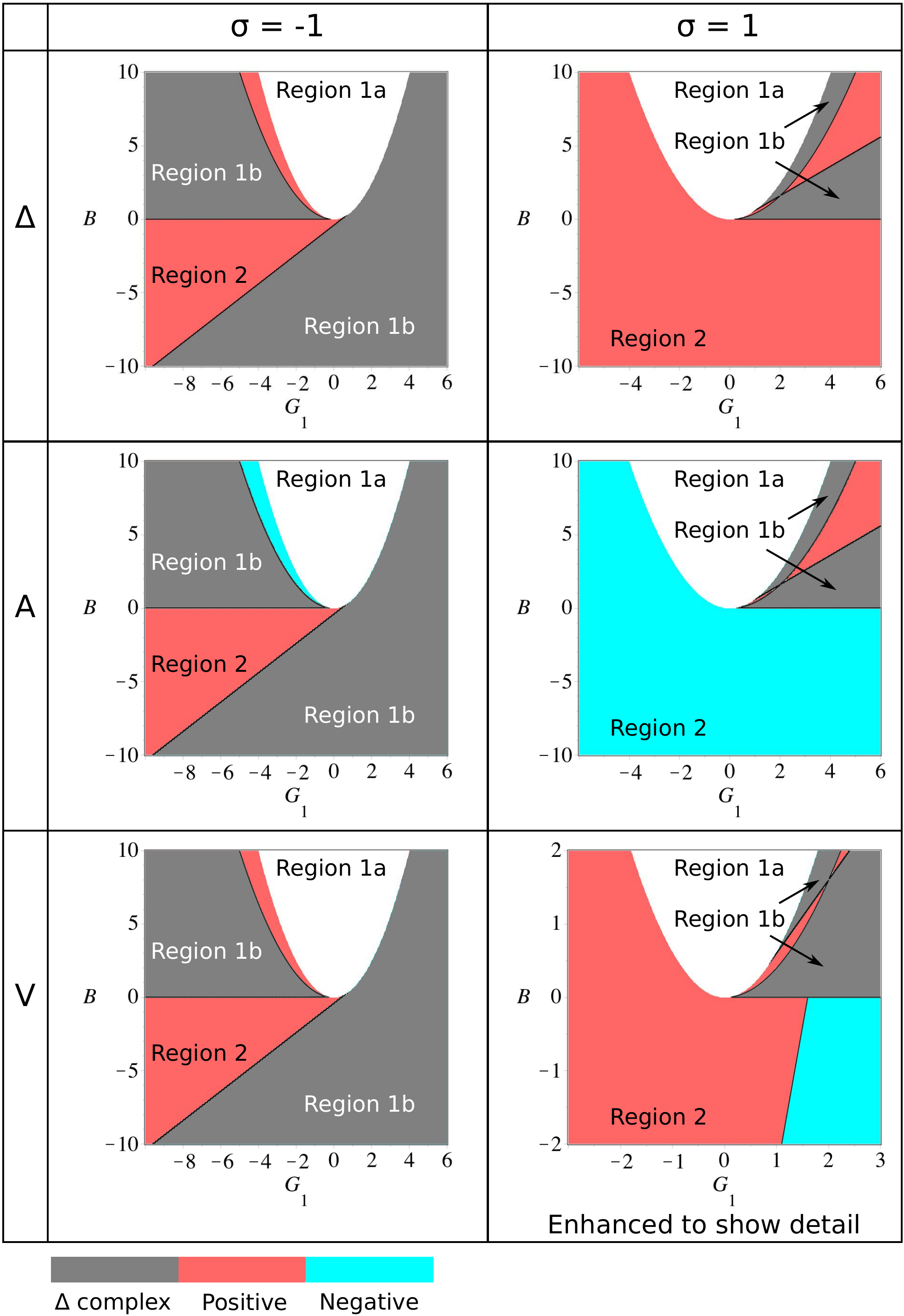}
    \caption{(color online) The parameter plane $(B, G_1)$ for the cases $\sigma = \pm 1$ in equation \eqref{SolHW13} where soliton solutions in the form of \eqref{TrialSol} can exist in region 2. Region 1a and region 1b denote areas where $\Delta$ is complex. Subsequent panels show subregions of region 2, where soliton speed or amplitude is positive (red) or negative (blue). The images for $\Delta$ use red shading for region 2, for reference. We use an enhanced scale in the lower right frame for $V$ to show detail.}
\label{fig:ESPlotsp1}
\end{figure}

In the case of $G_2 = 3G_1$ equation \eqref{SolHW} for the soliton half-width reduces to
\begin{equation}
\Delta_{1,2}^2 = \frac{80B \lb 2G_1^2 - 5Bs \rb}{5B \lb 3G_1 + 4s \rb - 20G_1^2 + \sigma \lb 5B - 4G_1 \rb \sqrt{5 \lb 5G_1^2 - 8Bs \rb}},
\label{SolHW13}
\end{equation}
where $\sigma = \pm 1$. We plot the parameter plane $(B, G_1)$ for the half-width, amplitude and speed in Figure \ref{fig:ESPlotsm1} for $s = -1$ and $\sigma = \pm 1$, and similarly in Figure \ref{fig:ESPlotsp1} for $s = 1$ and $\sigma = \pm 1$. Soliton solutions in the form of equation \eqref{TrialSol} cannot exist if $5G_1^2 - 8Bs < 0$, that is $B < -5 G_1^2/8$ for $s = -1$ and $B > 5 G_1^2/8$ for $s = 1$ (denoted as region 1a in Figures \ref{fig:ESPlotsm1} and \ref{fig:ESPlotsp1}) or if $\Delta_{1,2}^2 < 0$ (region 1b). The regions where $\Delta_{1,2}^2 > 0$ are designated by region 2. For the plot of soliton speed and amplitude, red regions indicate a positive quantity and blue regions indicate a  negative quantity (note that, in the figures for $\Delta$, we assume the positive square root is taken). The difference between the cases for $\sigma = -1$ and $\sigma = 1$ is shown in the first line of each figure, where the size of region 1b differs in each case and therefore the range of values for which solitons can exist is different in each case.

Thus, we see that nonlinear dispersion can affect the existence of
soliton solutions, their nature (regular or embedded solitons),
polarity, and, apparently, stability, as discussed in section
\ref{Sect3}.

\section{Numerical solutions for stationary solitons}
\label{Sect3}
Soliton solutions derived in the previous section represent particular cases of the wide family of stationary solutions containing a class of solitary waves. Solitary wave solutions of equation \eqref{KdV2K} (or in the stationary case equation \eqref{SolODE}) can be constructed, in general, by means of one of the well-known numerical methods, e.g., the Petviashvili method \cite{Petviashvili92, Pelinovsky04} or Yang--Lacoba method \cite{Yang08}. As mentioned in the Introduction, in some particular cases soliton solutions were found analytically, in other cases numerically. In the particular case of the Gardner--Kawahara equation, when $\gamma_1 = \gamma_2 = 0$ in equation \eqref{KdV2K} (or $G_1 = G_2 = 0$ in equation \eqref{SolODE}), soliton solutions were constructed and studied in \cite{Kurkina15}.

Here we will construct a family of soliton solutions using the Petviashvili method for some particular cases to compare the numerical solutions with the analytical solutions derived in the previous section. The numerical method in application to equation \eqref{SolODE} is described in Appendix \ref{AppendB}. We consider two particular cases when $G_1 =
G_2 = 0$ and when $G_2 = 3G_1$. In the latter case we will see the influence of nonlinear dispersion on the shape and polarity of solitary waves.

\subsection{The Gardner--Kawahara equation ($G_1 = G_2 = 0$)}
\label{Subsec3.1}%

We showed in Section \ref{Subsubsec2.1.1} that embedded solitons of the form \eqref{TrialSol} can exist in the Gardner--Kawahara equation under certain restrictions on the value of $B$, and these restrictions are different for $s = -1$ and $s = 1$. We therefore split the following analysis into two subsections.

\subsubsection{Numerical solutions in the case of negative cubic term ($s = -1$)}
\label{Subsubsec3.1.1}%
\begin{figure}[!htbp]%
\begin{center}%
\includegraphics[width=16cm]{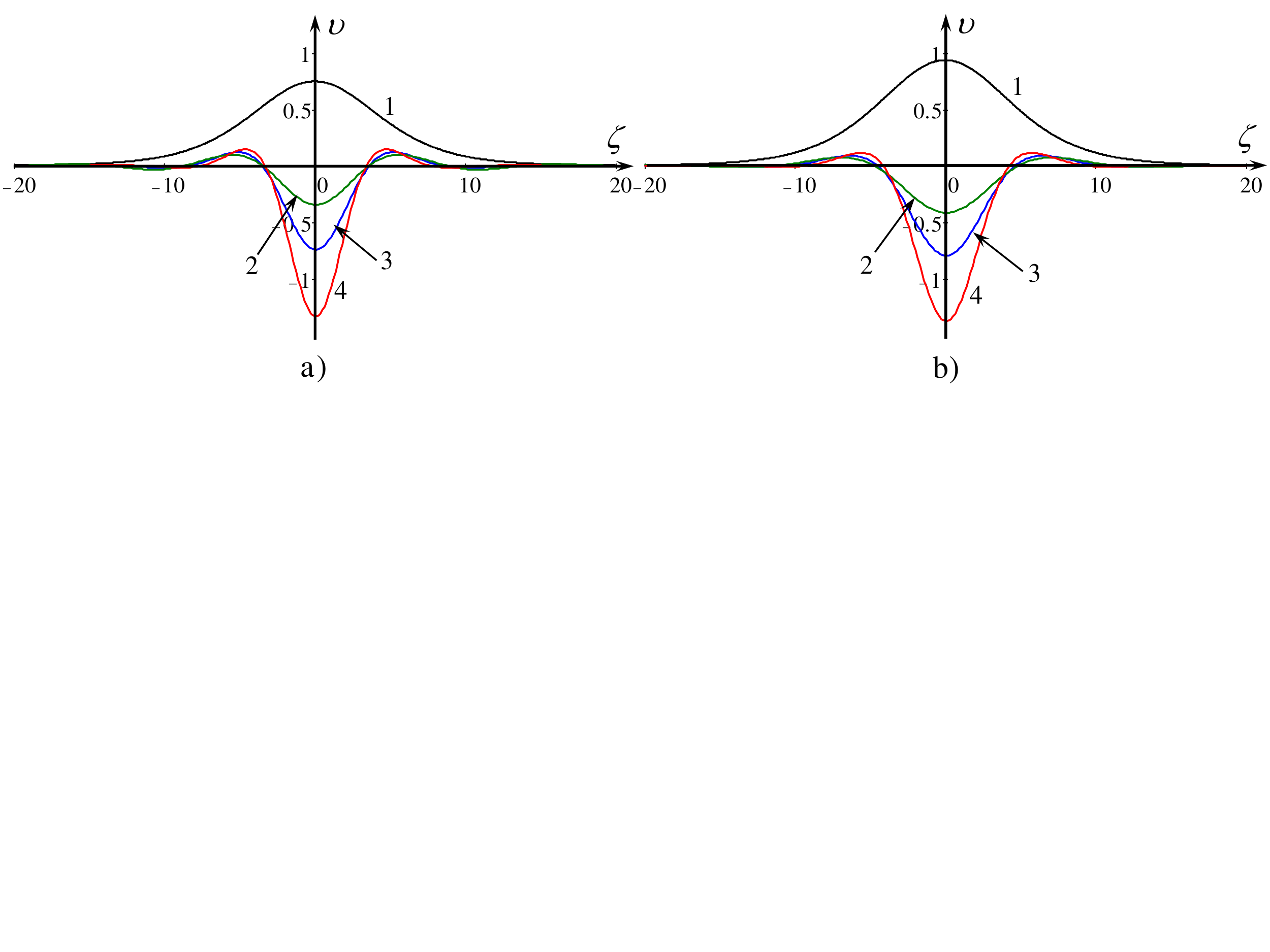} %
\end{center}
\vspace*{-7.5cm}%
\caption{(color online) Soliton profiles in the case $G_1 = G_2 =
0$. Panel (a) $B = 8/5$: the embedded soliton \eqref{TrialSol}
(line 1), other lines represent regular solitons for $V = -0.25$
(line 2), $V = -0.5$ (line 3), and $V = -1$ (line 4). Panel (b) $B
= 128/45$: the embedded soliton \eqref{TrialSol} (line 1), other
lines represent regular solitons for $V = -0.25$ (line 2), $V =
-0.5$ (line 3), and $V = -1$ (line
4).} %
\label{fig:GKsm1}
\end{figure}

In this case soliton solutions in the form of equation
\eqref{TrialSol} can exist only for $B > 2/5$ and represent the
embedded soliton. However, soliton solutions in different forms
can exist both for positive and negative $B$. For positive $B$
such solutions can be constructed numerically for $V < V_{min}
\equiv -1/(4B)$ (see Figure \ref{fig:PhaseSpeeds}). In Figure
\ref{fig:GKsm1} we present two families of numerical solutions. In
panel (a) one can see the analytical solution for the embedded
soliton \eqref{TrialSol} (line 1) when $B = 8/5$ and the soliton
has the minimum width $\Delta_{min} = \sqrt{32}$ and speed $V =
3/20$ (see Subsection \ref{Subsubsec2.1.1}). Lines 2, 3, and 4
show numerically constructed regular solitons for the same value
of $B$ and $V = -0.25$, $-0.5$, and $-1$ respectively.

In panel (b) one can see the analytical solution for the embedded
soliton \eqref{TrialSol} (line 1) when $B = 128/45$ and the
soliton has the maximal speed $V = 5/32$ and width $\Delta =
16\sqrt{2/15}$ (see Subsection \ref{Subsubsec2.1.1}). Lines 2, 3,
and 4 show numerically constructed regular solitons for the same
value of $B$ and $V = -0.25$, $-0.5$, and $-1$ respectively.

\begin{figure}[h!]%
\begin{center}%
\includegraphics[width=16cm]{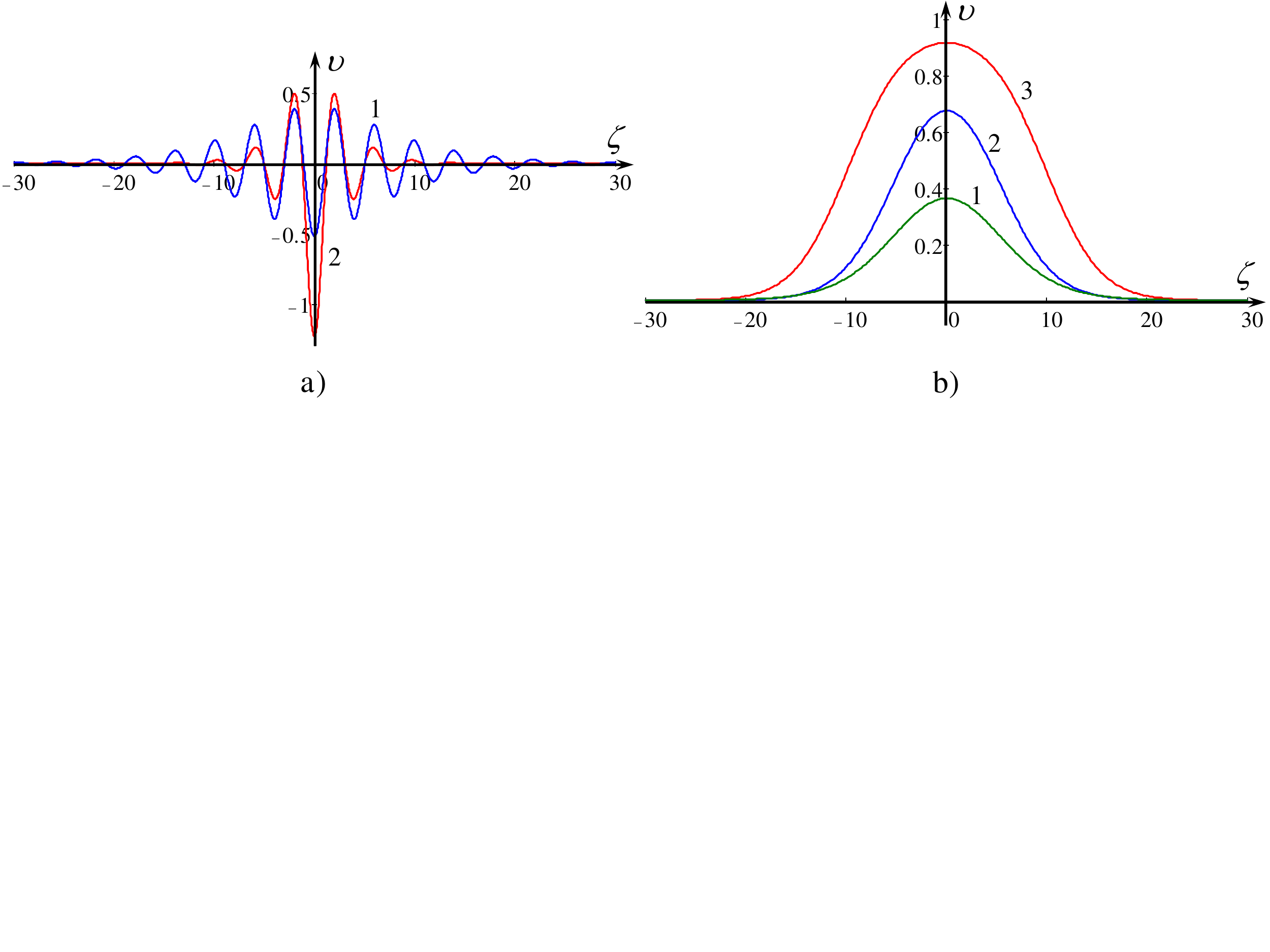} %
\end{center}
\vspace*{-7.5cm}%
\caption{(color online) Soliton profiles in the case $G_1 = G_2 = 0$. Panel (a) shows solitons for $B = 1/5$ and speed $V = -1.3$ (line 1) or $V = -1.6$ (line 2). Panel (b) shows solitons for $B = -1$ and speed $V = 0.1$ (line 1), $V = 0.15$ (line 2) or $V = 0.17$ (line 3).}
\label{fig:GKsm1Num}
\end{figure}

When $B < 2/5$ analytical solutions of equation \eqref{SolODE} are
unknown, however they can be constructed numerically. We managed
to construct such solutions within a relatively narrow range of
parameters. In particular, when $B = 1/5$, solitary-type solutions
appear in the form of oscillatory wave trains -- see Figure
\ref{fig:GKsm1Num}a (similar solutions were constructed
numerically in \cite{Champneys97}). When $V$ approaches $-1.25$
from below, solitons look like envelope solitons of the non-linear
Schr\"{o}dinger (NLS) equation (see, e.g. \cite{Whitham74,
Karpman75, Lamb80, Ablowitz81, Dodd82, Newell85}). Then, when $V$
decreases, the soliton shape smoothly changes and represents a
negative polarity soliton with oscillatory tails as shown by line
2 in Figure \ref{fig:GKsm1Num}a. We did not manage to construct
numerical solutions for $V < -1.7$.

When $B < 0$ regular solitons exist for $V > 0$ and have
bell-shaped profiles as shown in Figure \ref{fig:GKsm1Num}b for
$B = -1$ and several values of $V$. However, they exist only
within a relatively narrow range of $V$ between $0$ and $0.17$.
Similar situation occurs for other negative values of $B$.

\subsubsection{Numerical solutions in the case of positive cubic term ($s = 1$)}
\label{Subsubsec3.1.2}%

In this case solutions in the form of \eqref{TrialSol} can exist
only for negative $B$; they can be in the form of either regular
solitons or embedded solitons (see Subsection
\ref{Subsubsec2.1.1}). However soliton solutions in different
forms can exist, in principal, both for positive and negative $B$.
Nevertheless we did not manage to construct soliton solutions
numerically for positive $B$. This requires further investigation.

\begin{figure}[!htbp]%
\begin{center}%
\includegraphics[width=17cm]{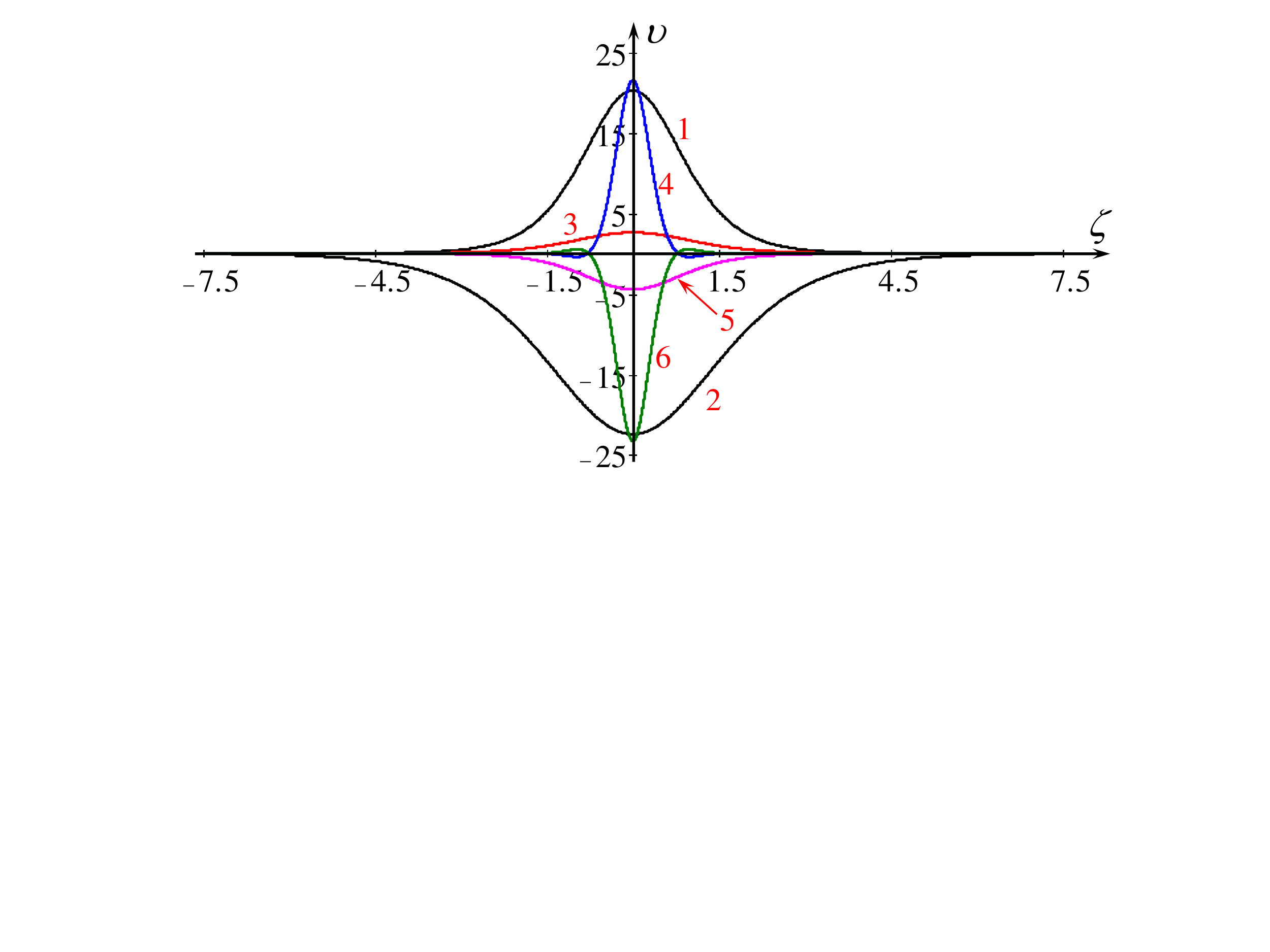} %
\end{center}
\vspace*{-7cm}%
\caption{(color online) Soliton profiles in the case of positive
cubic nonlinearity, $s = 1$, with $G_1 = G_2 = 0$ and $B = -0.1$.
Lines 1 and 2 show analytical solutions \eqref{TrialSol} for
regular solitons with $V = 2.1$ for the positive polarity soliton
and $V = 0.9$ for the negative polarity soliton. Lines 3 and 5
pertain to the numerical solutions with $V = 2.1$ and $V = 0.9$
correspondingly, and lines 4 and 6 pertain to solutions with $V =
100$.} \label{fig:GKsp1}
\end{figure}

In the case of negative $B$ there are two families of regular
solitons of positive and negative polarities. Typical examples are
shown in Figure \ref{fig:GKsp1} for $B = -0.1$. Line 1 represents
the analytical solution \eqref{TrialSol} with $V = 2.1$ and line 2
is another analytical solution with $V = 0.9$ as per equation
\eqref{VRestr}. However,  we failed to reproduce these solutions
numerically by means of the Petviashvili method. Lines 3 and 5 in
Figure \ref{fig:GKsp1} correspond to solutions found using the
numerical scheme for $V = 2.1$ and $V = 0.9$ respectively (in
comparison to the analytical solutions of line 1 and line 2). All
these solitons have exponentially decaying asymptotics at
infinity.

To explain the difference in the solutions, we refer to equation
\eqref{AlgEqForMu}. As follows from the analysis of this equation,
all its roots are real when $0 < V < -1/(4B)$ (which is the case
for Figure \ref{fig:GKsp1}). This means that exponential decay
of a solitary wave can be controlled by one of two real roots at
plus infinity and one of the other two roots at minus infinity.
Apparently, soliton solutions obtained analytically and
numerically correspond to different roots of characteristic
equation \eqref{AlgEqForMu}.

Then, it follows from numerical solutions that when the soliton
speed increases, so does the amplitude, but the soliton profile
becomes non-monotonic; this is again in agreement with the
qualitative analysis shown in Figure \ref{fig:RootSchem}a for $B
< 0$. Lines 4 and 6 illustrate such solutions for $V = 100$.

We did not manage to construct solitons of negative polarity with $B = -0.1$ and $V < 1.95$ even when the starting solution for the iteration scheme (see Appendix \ref{AppendB}) was chosen in the form of the analytical solution \eqref{TrialSol}. However, for $V \ge 1.95$, solitons of negative polarity were constructed numerically and are shown in Figure \ref{fig:GKsp1}. They represent almost mirror reflections of solitons of positive polarity. A similar situation occurs in the case of the Gardner equation with positive cubic nonlinearity, when solitons of positive polarity can exist for all amplitudes from 0 to infinity, whereas solitons of negative polarity cannot exist, if their amplitudes are less then some critical value. Below the critical amplitude breathers can exist instead (see, e.g., \cite{OstrEtAl15} and references therein).

\subsection{The particular case of equation \eqref{SolODE} with $G_2 = 3G_1$}
\label{Subsec3.2}%

Another particular case we study is $G_2 = 3G_1$. The basic
equation \eqref{SolODE} in this case contains four independent
parameters $B$, $G_1$, $s$, and $V$, which  determine the
structure of solitary waves. This case is convenient from the
numerical point of view as the iteration scheme is simpler than
for other cases of $G_1$ and $G_2$. Another point of interest in
this case is the fact that wave energy is not conserved in general
-- see equation \eqref{EnBal} (whereas for even solutions the
energy is conserved), therefore this allows us to understand the
role of nonlinear dispersion. In the parameter plane $(G_1,B)$
there are zones where regular solitons of different polarity can
exist. The boundaries of these zones are fairly complicated and we
demonstrate some typical soliton solutions belonging to different
zones, for the cases when $s = -1$ and $s = 1$ as before.

\subsubsection{Numerical solutions in the case of negative cubic term ($s = -1$)}
\label{Subsubsec3.2.1}%

\begin{figure}[!htbp]%
\begin{center}%
\includegraphics[width=16cm]{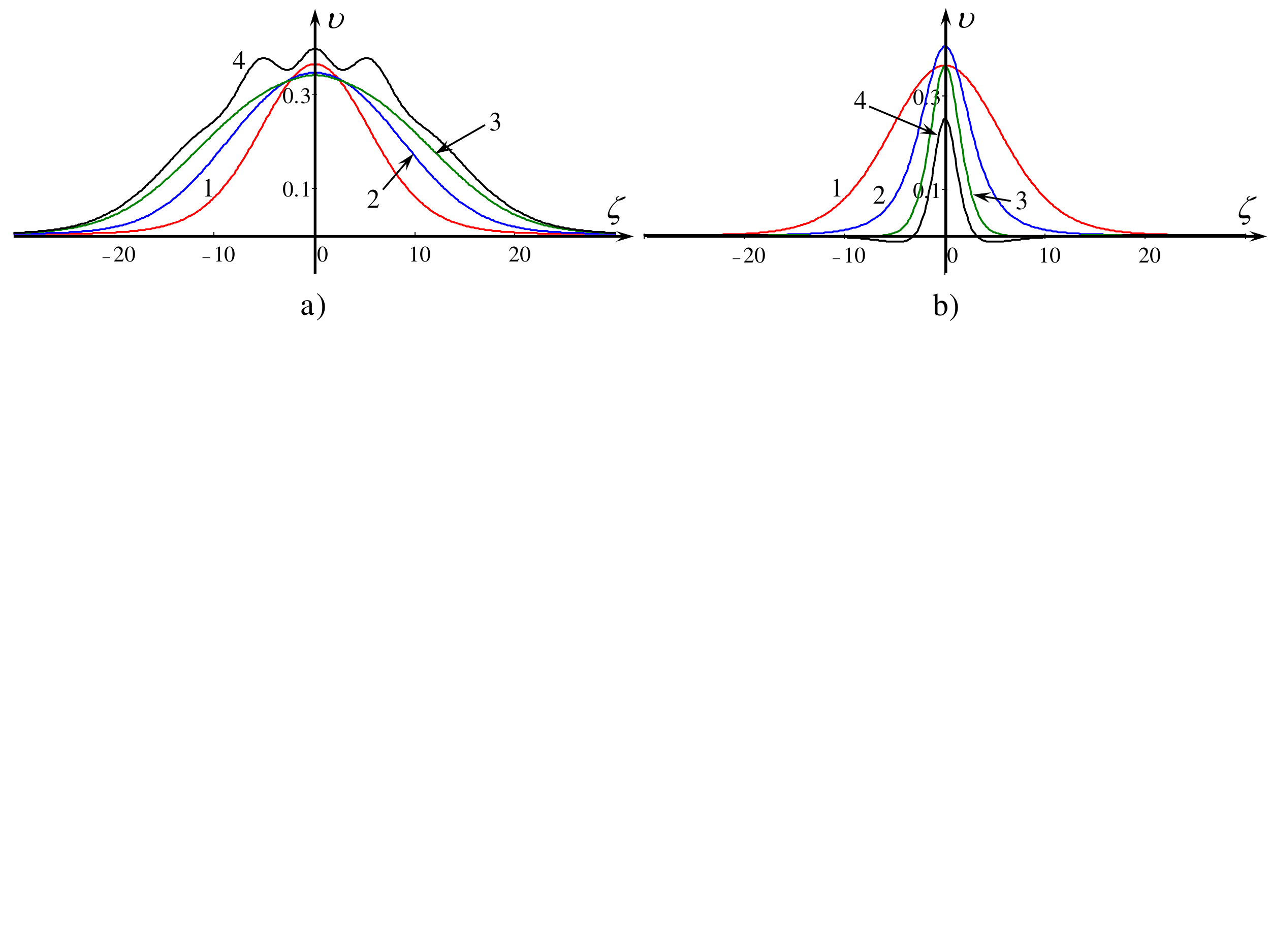} %
\end{center}
\vspace*{-8.5cm}%
\caption{(color online) Numerical solutions of equation \eqref{SolODE} for negative cubic nonlinearity ($s = -1$), $B = -1$, $V = 0.1$, $G_2 = 3G_1$, and several values of $G_1$. In
panel (a) we have the soliton solution for $G_1 = 0$ (line 1, reference case), $G_1 = -4$ (line 2), $G_1 = -8$ (line 3) and $G_1 = -9$ (line 4). In panel (b) we have the soliton solution for $G_1 = 0$ (line 1, reference case), $G_1 = 4$ (line 2), $G_1 = 8$ (line 3) and $G_1 = 16$ (line 4).}%
\label{fig:G23G1sm1Bm1}
\end{figure}

\begin{itemize}
\item $B = -1$ %
\end{itemize}
Consider first the case when $B = -1$ in equation \eqref{SolODE}.
In this case regular solitons can exist for $V > 0$. In Figure
\ref{fig:G23G1sm1Bm1} we illustrate the structure of a solitary
wave when $G_1$ varies. Line 1 in both panels show the reference
case when $G_1 = 0$. If this parameter becomes negative, the
solitons become wider, their amplitudes slightly decrease, and the
tops become flatter (see lines 2 and 3 in panel a). However, when
$G_1$ approaches $-9$, the soliton profiles become non-monotonic
with oscillations on the top (see line 4 in panel a) (similar
solutions were constructed numerically in
\cite{Champneys97}). For smaller values of $G_1 < -9$ we were
unable to obtain numerical solutions (apparently, they do not
exist for such a set of parameters).

When $G_1$ becomes positive and increases the solitons become narrower, their amplitudes slightly increase first, but then they monotonically decrease when $G_1$ exceeds 4 (see lines 1, 2, and 3 in panel b). For relatively large values of $G_1$ soliton tails become non-monotonic with negative minima on the profile (see line 4 in panel b).

\begin{figure}[!htbp]%
\begin{center}%
\includegraphics[width=16cm]{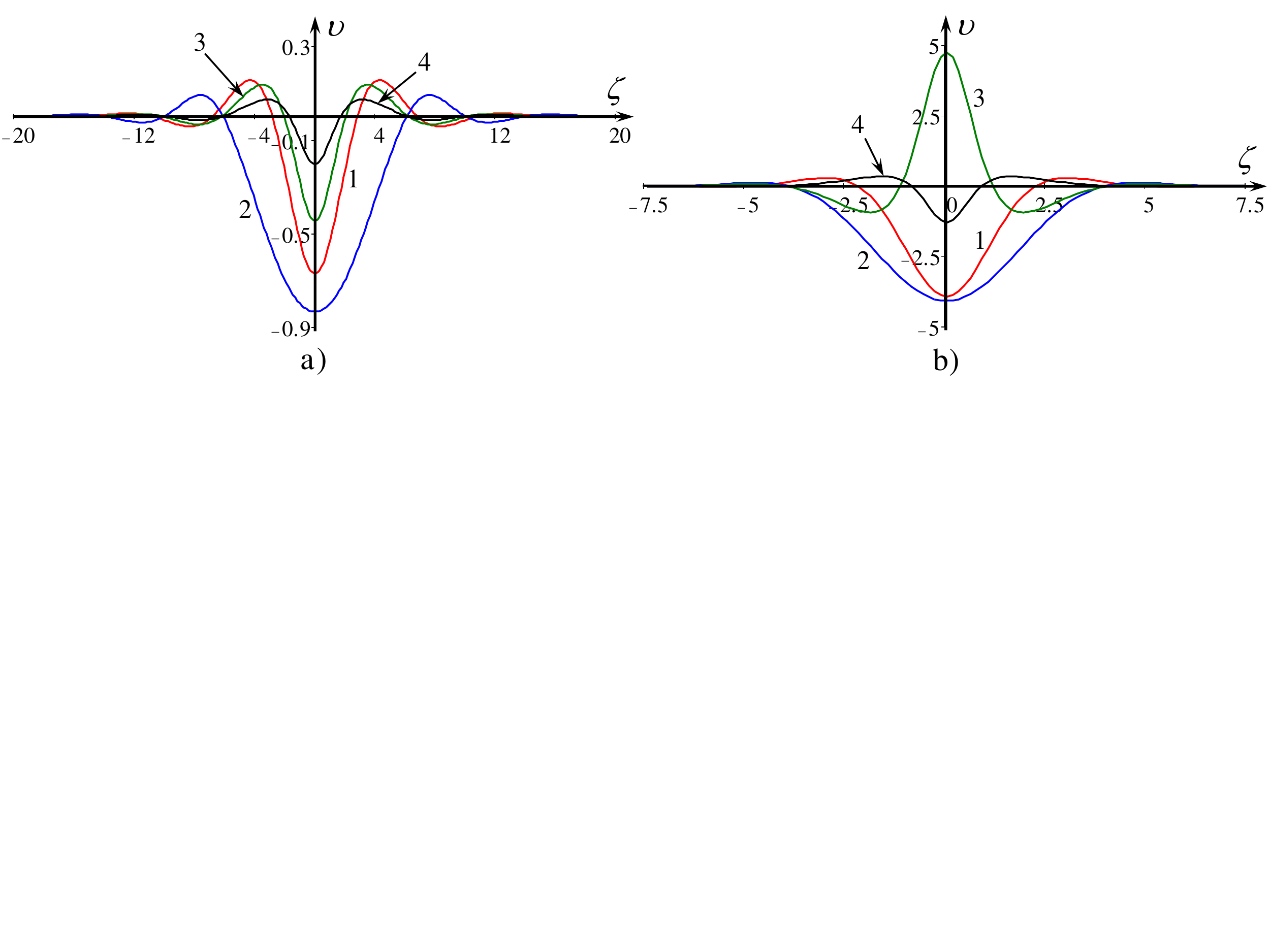} %
\end{center}
\vspace*{-7.5cm}%
\caption{(color online) Numerical solutions of equation \eqref{SolODE} for negative cubic nonlinearity ($s = -1$), $B = 1$, $G_2 = 3G_1$, and several values of $G_1$. Panel (a): $V = -0.5$ and we have the soliton solution for $G_1 = 0$ (line 1, reference case), $G_1 = -3$ (line 2), $G_1 = 3$ (line 3) and $G_1 = 12$ (line 4). Panel (b): $V = -5$ and we have the soliton solution for $G_1 = 0$ (line 1, reference case), $G_1 = -1.88$ (line 2), $G_1 = -1.89$ (line 3) and $G_1 = 12$ (line 4).}
\label{fig:G23G1sm1Bp1}
\end{figure}

\begin{itemize}
\item $B = 1$ %
\end{itemize}
When $B=1$ solutions in the form of regular solitons can exist only
for negative $V < V_{min} \equiv -1/(4B)$ (see after equation
\eqref{DispRel}). In Figure \ref{fig:G23G1sm1Bp1}a we show the
structure of solitary waves for $V = -0.5$ and different values of
$G_1$. Solitons in this case have negative polarity and
oscillating tails. Line 1 corresponds to the case when $G_1 = 0$.
If this parameter becomes negative, the solitons become wider and
their amplitudes increase (see line 2 in panel a). For $G_1 < -3$
we were unable to construct numerical solutions (apparently, they
do not exist for such a set of parameters). When $G_1$ becomes
positive and increases, the solitons become narrower and smaller
(see lines 3 and 4 in panel a).

In Figure \ref{fig:G23G1sm1Bp1}b we show the structure of solitary waves for $V = -5$ and different values of $G_1$. Solitons in this case can have both negative and positive polarity; they can have slightly oscillating tails or non-monotonic aperiodic tails. Line 1 corresponds to the case when $G_1 = 0$. If this parameter becomes negative and varies from zero to $G_1 = -1.88$, the solitons remain qualitatively the same, but become wider and their amplitudes slightly increase as shown in panel b) by line 2. When $G_1$ further decreases and becomes less than $-1.89$, the solitons abruptly change their polarity, become taller and narrower with well-pronounced negative minima (see line 3). Further increases in $G_1$ result in soliton profiles that remain qualitatively the same, but their amplitudes decrease.

When $G_1$ becomes positive and increases, soliton profiles remain similar to the shape of the soliton for $G_1 = 0$, but the solitons become narrower and of smaller amplitude (see, e.g., line 4 in panel b).

\subsubsection{Numerical solutions in the case of positive cubic term ($s = 1$)}
\label{Subsubsec3.2.2}%

In this case the analytical solution \eqref{TrialSol} for $B > 0$
can exist only for $V > 0$ representing an embedded soliton (see
Figure \ref{fig:PhaseSpeeds} and Figure \ref{fig:ESPlotsp1}),
whereas solutions in the form of regular solitons, apparently,
cannot exist; we were unable to construct such solutions
numerically.

\begin{figure}[!htbp]%
\begin{center}%
\includegraphics[width=16cm]{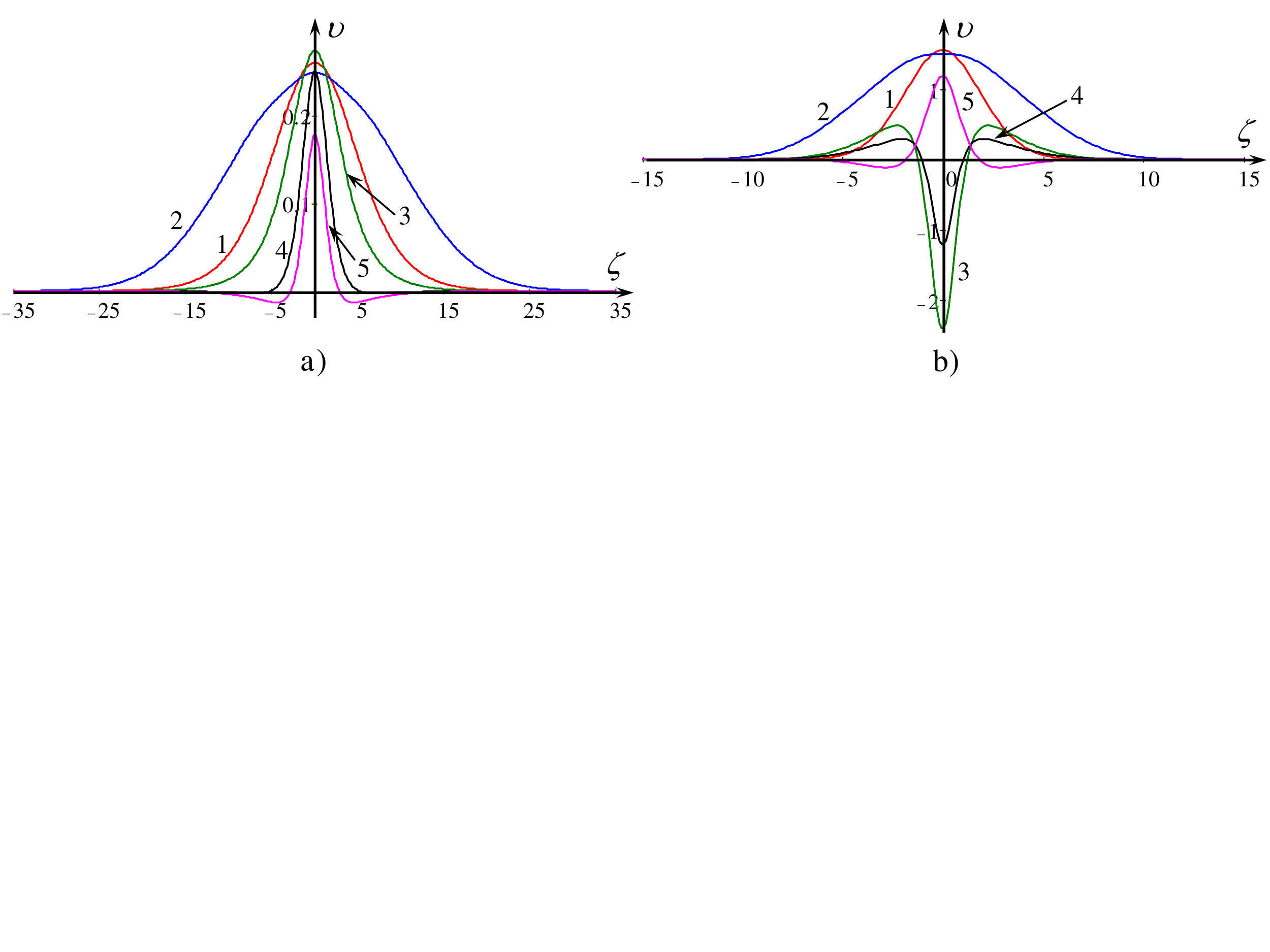} %
\end{center}
\vspace*{-7.5cm}%
\caption{(color online) Numerical solutions of equation \eqref{SolODE} for positive cubic nonlinearity ($s = 1$) and $G_2 = 3G_1$, $B = -1$. Panel (a): $V = 0.1$ and we have the soliton solution for $G_1 = 0$ (line 1, reference case), $G_1 = -12$ (line 2), $G_1 = 4$ (line 3), $G_1 = 12$ (line 4) and $G_1 = 24$ (line 5). Panel (b): $V = 1$ and we have the soliton solution for $G_1 = 0$ (line 1, reference case), $G_1 = -4.63$ (line 2), $G_1 = -4.64$ (line 3), $G_1 = -12$ (line 4) and $G_1 = 6$ (line 5).}%
\label{fig:G23G1sp1}
\end{figure}

For $B < 0$ the analytical solution \eqref{TrialSol} exists in the
form of a regular soliton for $V > 0$ and embedded soliton for $V
< 0$ (see Figure \ref{fig:PhaseSpeeds} and Figure
\ref{fig:ESPlotsp1}). Embedded solitons cannot be reproduced by
means of the Petviashvili method (see Appendix \ref{AppendB}),
therefore we constructed only regular solitons numerically for the
particular case of $B = -1$ and $V > 0$. The structure of these
solitons depends on their speed and is illustrated by Figure
\ref{fig:G23G1sp1} for two particular values of $V$. In panel (a)
we show a few typical soliton profiles for $V = 0.1$ and several
values of $G_1$. Line 1 shows the reference case when $G_1 = 0$.
If $G_1$ becomes negative, the solitons become wider and their
amplitudes slightly decrease (see line 2). For $G_1 < -12$ we did
not obtain soliton solutions.

When $G_1$ becomes positive, the soliton amplitude slightly
increases first (see line 3), but then it decreases, and solitons
become narrower (see line 4). For sufficiently large $G_1$
solitons profiles become non-monotonic, so that negative minima
appear in the profiles (see line 5).

In panel (b) we show other typical soliton profiles for $V = 1$ and several values of $G_1$. Line 1 shows the reference case when $G_1 = 0$. If $G_1$ becomes negative, the solitons become wider and their amplitudes slightly decrease (see line 2). When $G_1$ passes through some critical value between $-4.63$ and $-4.64$ the soliton polarity abruptly alters from positive to negative (see line 3). Then, when $G_1$ further decreases, soliton profiles remain qualitatively similar to what is shown by line 3, but their amplitudes become smaller (see line 4). When $G_1$ becomes positive soliton profiles remain qualitatively similar to line 1, but their amplitudes gradually decrease and well-pronounced minima appear on both sides of the crests (see line 5).

The solutions constructed numerically do not reproduce the analytical solution \eqref{TrialSol}. The reason is the same as discussed in Section \ref{Subsubsec3.1.2} i.e.  solutions with different asymptotics can coexist for the same set of parameters, and the numerical scheme, apparently, converges to only one of them which is different from solution \eqref{TrialSol}.

\subsection{Stationary solutions for two particular cases of equation \eqref{DivForm}}
\label{Subsec3.3}%
\begin{figure}[h!]
\includegraphics[width = 10cm]{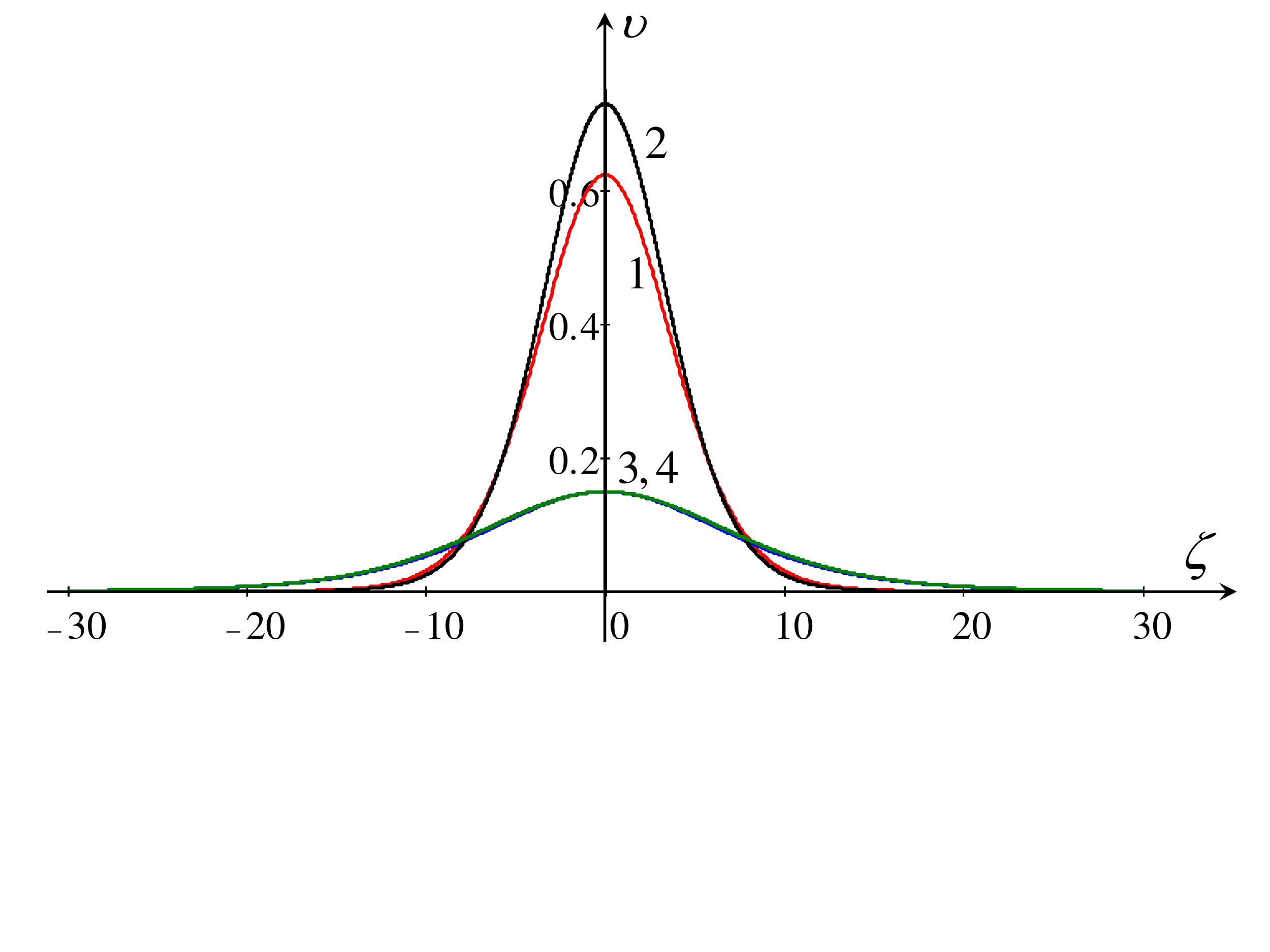}
\vspace*{-2.5cm}%
\caption{(color online) Yamamoto--Takizawa soliton \eqref{YT-Soliton} (line 1). Other lines represent numerically obtained soliton solutions of equation \eqref{DivForm} with $s = G_1 = G_2 = 0$ and $B = -1$. The solution with $V = 0.25$ and $V = 0.15$ correspond to line 2 and line 3 respectively, and line 4 represents the KdV soliton with the same amplitude as line 3.}
\label{fig:YT-sol}%
\end{figure}

In this subsection we briefly consider particular exact solutions of equation \eqref{DivForm}. In the first case we set $s = G_1 = G_2 = 0$ to reduce the general equation \eqref{DivForm} to the generalised Kawahara equation containing both third- and fifth-order derivatives. The exact solution to this equation was obtained for the first time by  Yamamoto \& Takizawa \cite{Yamamoto81} (for further references see also \cite{Kichenassamy92})
\begin{equation}
\upsilon \lb \zeta \rb = -\frac{105}{169B} \sechn{4}{\frac{x - V \tau}{\Delta}},
\label{YT-Soliton}
\end{equation}
where $V = -36/169B$, $\Delta = \sqrt{-52B}$ and $B < 0$. There
are no free parameters in this solution; the amplitude, speed and
width of Yamamoto--Takizawa (YT) soliton \eqref{YT-Soliton} are
determined by the coefficients of the generalised Kawahara
equation. The soliton moves with a positive speed, whereas linear
waves propagate with negative phase speeds, therefore it is a
regular soliton; its profile is shown in Figure \ref{fig:YT-sol}
by line 1 for $B = -1$. It was easily reproduced numerically with
the help of the Petviashvili method when the speed was chosen in
accordance with the formula $V = -36/169$, $B \approx 0.213$.

By means of Petviashvili's method we constructed a family of
soliton solutions with the fixed value of parameter $B = -1$. All
solitary solutions of this family are qualitatively similar to the
YT soliton. In particular, line 2 shows the numerical solution for
$V = 0.25$, and line 3 for $V = 0.15$. It was discovered that the
soliton amplitude decreases as the speed decreases. At small
amplitudes the soliton profile becomes indistinguishable from the
profile of the KdV sech$^2$-soliton of the same amplitude. This is
illustrated by Figure \ref{fig:YT-sol} where the numerically
obtained line 3 practically coincides with line 4, which
represents the KdV soliton of the same amplitude. Apparently
within this equation there is a continuous family of solitary wave
solutions whose profiles depend on their amplitude, and the YT
soliton is just one particular of the representatives of this
family.

Because the generalised Kawahara equation is non-integrable, one
can expect that soliton interactions are inelastic, i.e. in the
process of soliton collisions they radiate small-amplitude
trailing waves and, as a result, change their parameters. This
will be confirmed in Section \ref{Sect4}.

In the second case we present the soliton solution to the
Kaup--Kupershmidt equation \cite{Dodd82, Newell85,
Kichenassamy92}. This equation is a particular case of equation
\eqref{KdV2K} with the following coefficients: $\alpha = \beta =
0$, $\varepsilon \alpha_1 = 180$, $\varepsilon \gamma_1 = 30$,
$\varepsilon \gamma_2 = 75$, and $\varepsilon \beta_1 = 1$. With
such a set of coefficients the equation is completely integrable,
and its soliton solution has a slightly unusual form:

\begin{equation}
\upsilon(\xi, \tau) = 3A \frac{4 \sechn{2}{\zeta / \Delta} -
\sechn{4}{\zeta / \Delta}}{\lsq 2 +  \sechn{2}{\zeta / \Delta}
\rsq^2},
\label{KKSoliton} %
\end{equation}
where $A$ is the soliton amplitude (a free parameter), $\Delta = 1/\sqrt{6A}$ is the soliton width, $\zeta = \xi - V \tau$, and $V = (6A)^2$ is the soliton speed.

\begin{figure}[h!]
\includegraphics[width = 10cm]{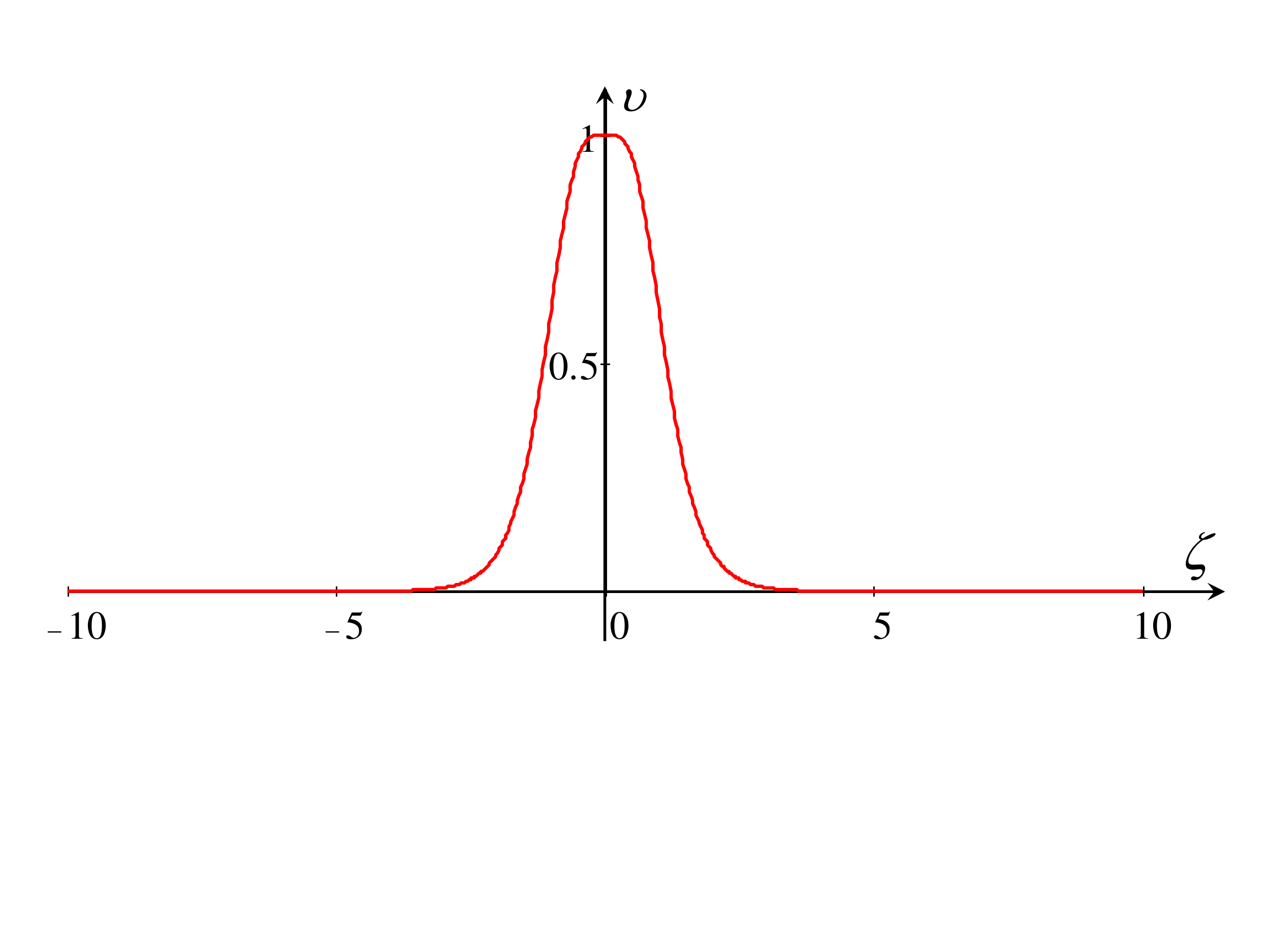}
\vspace*{-2.5cm}%
\caption{(color online) The Kaup--Kupershmidt ``fat'' soliton \eqref{KKSoliton} of a unit amplitude.}
\label{fig:KK-sol}%
\end{figure}

The profile of the Kaup--Kupershmidt (KK) soliton of a unit
amplitude is shown in Figure \ref{fig:KK-sol}. As one can see, its
speed is positive, whereas waves of infinitesimal  amplitude
within the Kaup--Kupershmidt equation have negative phase speeds
(in the moving coordinate frame). Therefore, the KK soliton is a
regular soliton too.

In contrast to the previous case, the Kaup--Kupershmidt equation
is completely integrable, therefore soliton interactions are
elastic, and to a certain extant, trivial. This means that after
interaction solitons completely restore their original parameters
and do not radiate small amplitude perturbations. Therefore,
solitons remain the same as they were before interaction, and the
only traces of interaction are their shifts in space and time,
exactly as in the interaction of elastic particles.

\section{Numerical study of soliton interactions}
\label{Sect4}%
It was anticipated in \cite{Kichenassamy92} that equation \eqref{KdV2K} would be studied numerically to confirm existence and robustness of solitary wave solutions, investigate whether they emerge evolutionary from the arbitrary initial pulse-type perturbations, how do they behave under collisions, whether ``they emerge unscathed as true solitons, or is there a small, but nonzero nonelastic effect''. Since that time such investigation was not carried out to the best of our knowledge. Here we will try to illuminate these issues.

To solve equation \eqref{KdV2K} numerically, we apply a
pseudospectral technique similar to that used in \cite{Grimshaw08,
Alias13, Alias14, Khusnutdinova17}. The equation is solved in the
Fourier space using a \nth{4} order Runge--Kutta method for time
stepping, while the nonlinear terms are calculated in the real
space and transformed back to the Fourier space for use in the
Runge--Kutta scheme. To remove the aliasing effects, we use the
truncation 2/3-rule by Orszag in Boyd \cite{Boyd01}. See Appendix
\ref{AppendC} for the description of the numerical scheme.

To generate solitons for a given set of parameters, an initial
pulse was taken as the initial condition. This pulse was taken in
the form a standard KdV soliton i.e. of the form
\begin{equation}
u(x,0) = P \sechn{2}{\sqrt{\frac{\abs{P}}{12}}\frac{x}{L}},
\label{Pulse}
\end{equation}
where $P$ corresponds to the amplitude and $L$ is a factor used to
change the width of this initial pulse. In each case, we can
control the number of solitons produced, and their amplitudes, via
the parameters $P$ and $L$.

To calculate the interaction of regular or embedded solitons, we
generate regular solitons of the required amplitude using the
method described above. Once the solitons generated from the pulse
are well separated, we extract them from the solution. To study
the interaction of these regular solitons with other regular
solitons or embedded solitons, we generate an initial condition
using the extracted soliton and either another extracted soliton
(for the interaction of regular solitons) or with the embedded
soliton found analytically, so that their interactions could be
studied. This extraction was performed so that any radiation
emerging from the initial pulse would not interfere with the
collision. In each of the numerical cases considered below, the
value of $P$ and $L$ are stated for each regular soliton generated
via this method. Furthermore, we state if the solitons used in the
proceeding calculations are found analytically or generated from a
pulse.

With the help of this numerical method we studied interactions of solitary waves with different parameters and different coefficients of the governing equation \eqref{DivForm}. First of
all we found that the embedded soliton propagates in all cases, with minimal loss of energy. We have calculated the change of ``wave energy'' $I_2 = \int (u^2/2) \dd{x}$ (see Introduction) as
\begin{equation}
\Delta I_2 = \frac{I_2(t) - I_2(0)}{I_2(0)},
\label{WaveEnergyChange}
\end{equation}
where $I_2(0)$ is the initial wave energy and $I_2(t)$ is the wave energy at time $t$.

Using this numerical method with periodic boundary conditions, we
obtained $\Delta I_2 = 1.2 \cdot 10^{-13}$ and $2.4 \cdot
10^{-13}$ for the embedded solitons in the cases when nonlinear
dispersion is present, whereas for the regular solitons in these
cases we obtained $\Delta I_2 = 1.2 \cdot 10^{-5}$ and $5.2 \cdot
10^{-6}$. In the case of the regular solitons, as they were
generated by a pulse-like initial condition, fast moving radiation
was generated that would re-enter the domain and interfere with
the main wave structure. To diminish this effect, we applied ``a
sponge layer'' to the solution domain to absorb this radiation and
prevent it re-entering the solution domain, as detailed in
Appendix \ref{AppendC}. This accounts for the lower accuracy in
the wave energy conservation. It is worth noting that when the
embedded soliton is perturbed, the energy is no longer conserved
and therefore the solution eventually breaks down, except in the
case when there is no nonlinear dispersion (see Section
\ref{Subsec4.1} below).

For the regular solitons in all cases, they steadily propagate
without loss of energy even in the cases when $G_2 \ne 2G_1$. As
has been mentioned above (see the paragraph after equation
\eqref{EnBal}), in the case of propagation of a stationary wave
described by an even function, the right-hand side of equation
\eqref{EnBal} vanishes, and wave energy $I_2$ is conserved.
However, an interesting question arises about the energy
conservation in the process of soliton interaction when
$G_2 \ne 2G_1$. Below we present the results of our numerical
study of equation \eqref{DivForm} with negative cubic
nonlinearity in the cases when (i) $G_1 = G_2 = 0$ and
(ii) $G_2 = 3G_1$. Equation \eqref{DivForm} with positive cubic
nonlinearity can be studied in a similar way, but such exercises
require much more computational resources, because soliton
amplitudes are limited in this case and can be very close to each
other, whereas their speeds are relatively small, therefore
soliton interactions take a very long time to compute.

\subsection{The Gardner--Kawahara equation ($G_1 = G_2 = 0$)}
\label{Subsec4.1}%

\begin{figure}[h!]
\includegraphics[width = 10cm]{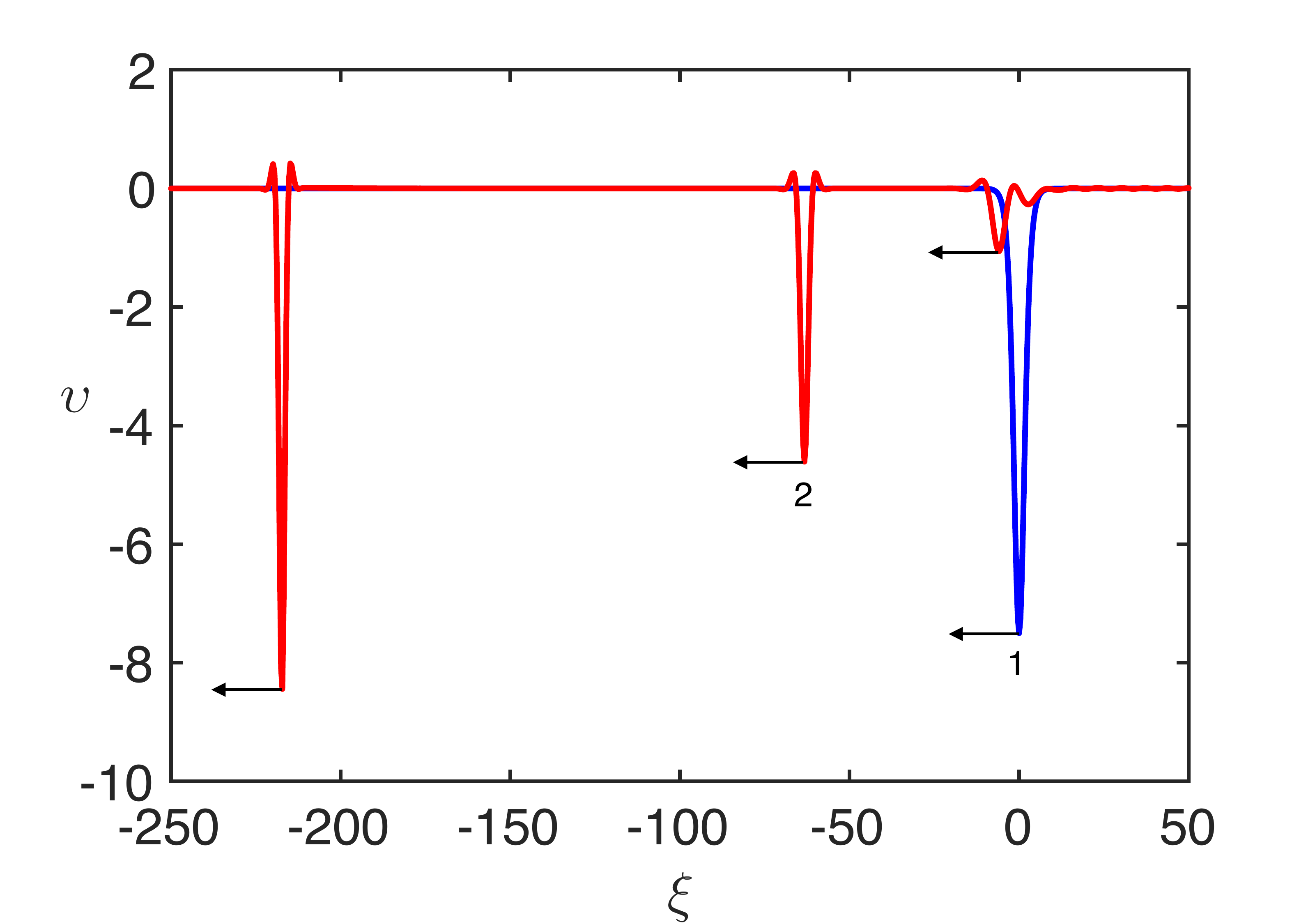}
\caption{(color online) Generation of several regular solitons from the initial sech$^2$-pulse within the framework of equation \eqref{DivForm} with $s = -1$, $B = 8/5$, $G_1 = G_2 = 0$. Line 1 corresponds to the initial condition at $t = 0$ (blue) and line 2 is the solution at $t = 10$ (red). The initial pulse parameters are $P = -7.5$ and $L = 1$.}%
\label{fig:5Ax5}%
\end{figure}

\begin{figure}[h]
\includegraphics[width = 10cm]{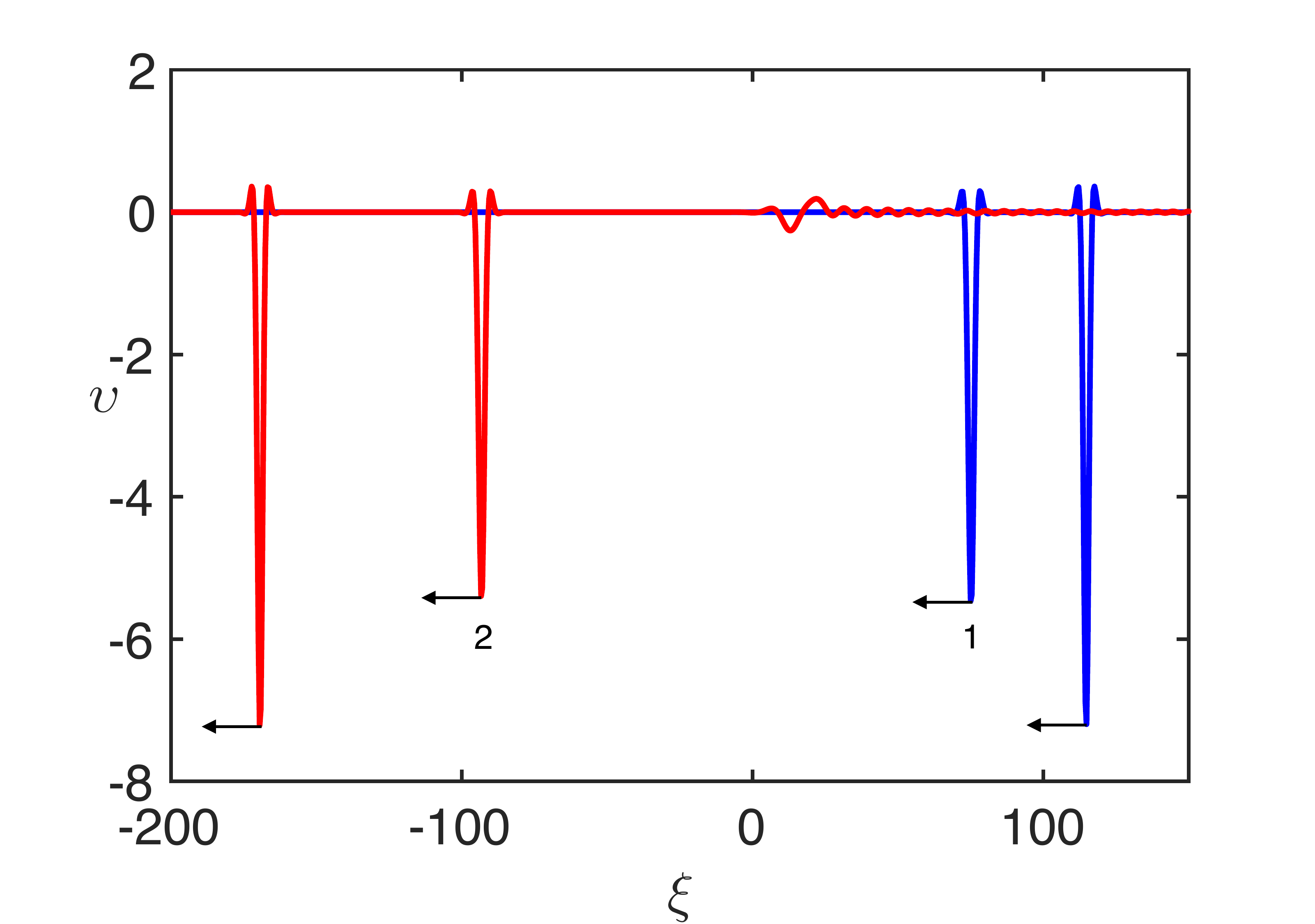}
\caption{(color online) Interaction of two regular solitons within
the framework of equation \eqref{DivForm} with $s = -1$, $B =
8/5$, $G_1 = G_2 = 0$. Line 1 corresponds to the initial condition
at $t = 0$ (blue) before the interaction and line 2 to the
solution after the interaction at $t = 20$ (red). Both solitons
were obtained numerically from a pulse-like initial condition. The
initial pulse parameters are $P_1 = -7.5$ and $L_1 = 1$ for the
taller soliton and $P_2 = 3P_1/4$ and $L_2 = 1$ for the shorter
soliton.}
\label{fig:5ACol}%
\end{figure}

\begin{figure}[h!]
\includegraphics[width = 10cm]{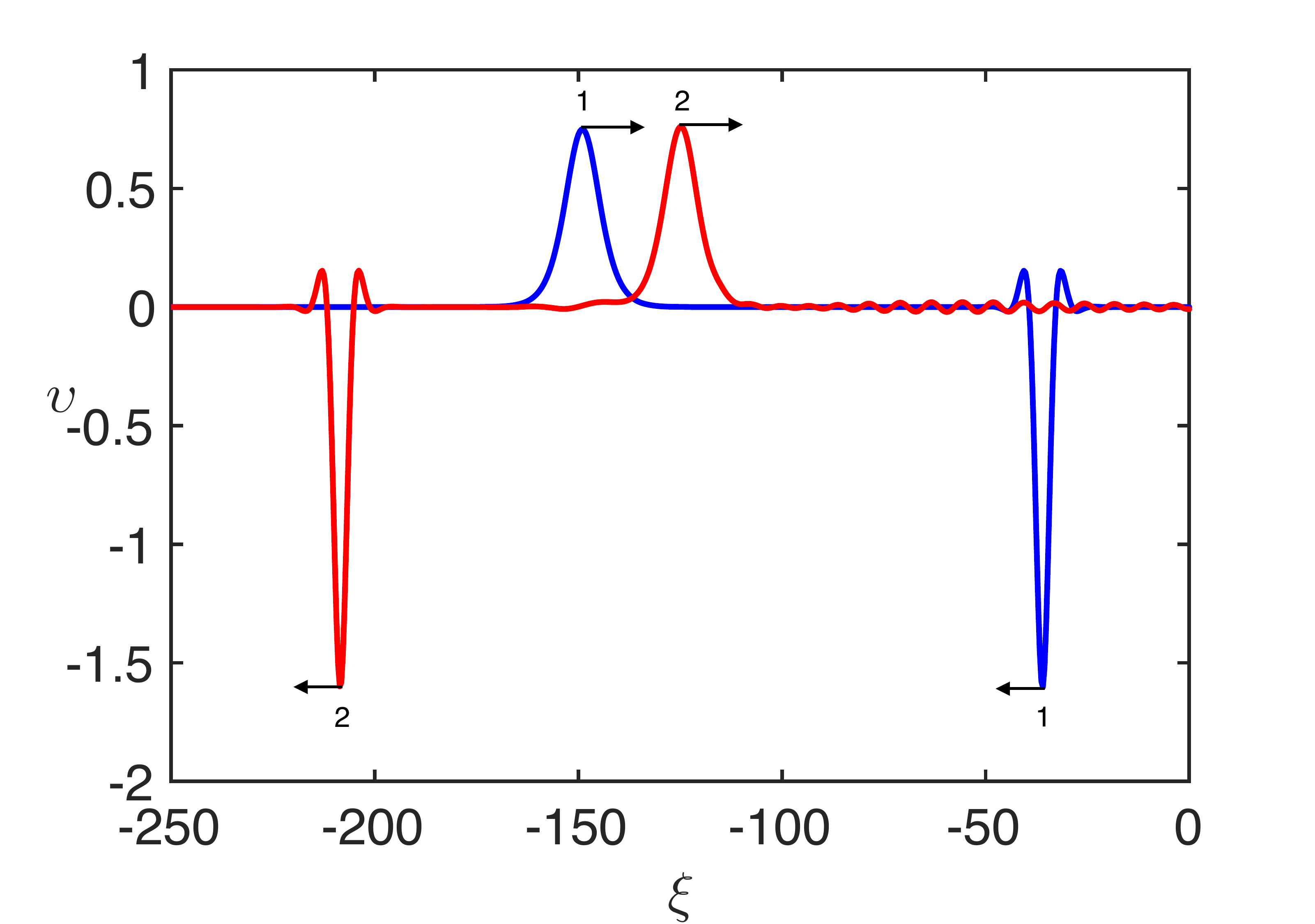}
\caption{(color online) Interaction of a regular soliton of
negative polarity and an embedded soliton of positive polarity
within the framework of equation \eqref{DivForm} with $s = -1$, $B
= -8/5$, $G_1 = G_2 = 0$. Line 1 corresponds to the initial
condition at $t = 0$ (blue) before the interaction and line 2 to
the solution after the interaction at $t = 140$ (red). The regular
soliton was obtained numerically from a pulse-like initial
condition. The initial pulse parameters are $P = -7.5$ and $L =
1$.}
\label{fig:5AColES}%
\end{figure}

As shown in subsection \ref{Subsubsec3.1.1}, there are families of
regular solitons for positive and negative $B$, some of which can
co-exist with the embedded soliton \eqref{TrialSol} -- see Figures
\ref{fig:GKsm1} and \ref{fig:GKsm1Num}. Here we present (i) an
example of pulse disintegration into a number of regular solitons
(Figure \ref{fig:5Ax5}); (ii) interaction of regular solitons
(Figure \ref{fig:5ACol}); and (iii) interaction of regular and
embedded solitons (Figure \ref{fig:5AColES}). Figure
\ref{fig:5Ax5} illustrates that regular solitons with
non-monotonic profiles asymptotically appear from pulse-type
initial perturbations in the process of its disintegration.
Apparently, such solitons can form bound states, the bi-solitons,
triple-solitons, or multi-solitons, i.e., stationary moving
formations consisting of two or more binding solitons
\cite{Gorshkov76, Gorshkov79, Champneys97, Obregon98, Kurkina15},
but we did not study this phenomenon in our paper.

The interaction of two solitons as shown in Figure \ref{fig:5ACol} demonstrates that the solitons survive after the collision, but a residual small wave packet is generated in the trailing wave field. This clearly indicates that the soliton collision is inelastic.

The most fascinating is Figure \ref{fig:5AColES} which
demonstrates (seemingly for the first time) that the embedded
soliton can survive after  interaction with a regular soliton. The
interaction is obviously inelastic, so that small disturbances
appear both in front and behind the embedded soliton. Thus, we see
that it survives even after collision with a regular soliton, and
for much longer times (up to $t = 700$) the embedded soliton keeps
its identity.

\subsection{Interactions of solitary waves when $G_2 = 3G_1$}
\label{Subsec4.2}%

\begin{figure}
\includegraphics[width = 10cm]{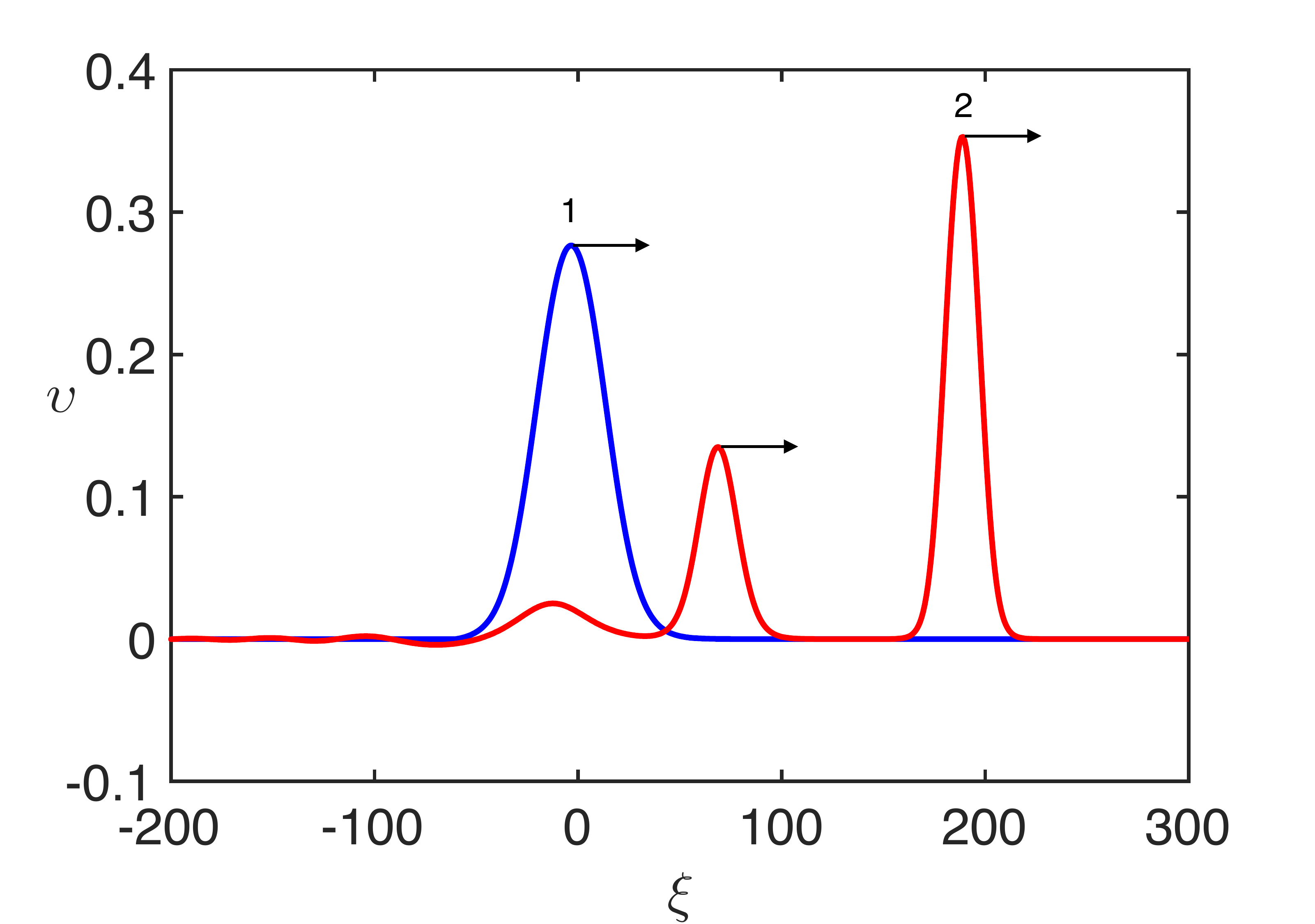}
\caption{(color online) Generation of regular solitons from the
initial sech$^2$-pulse within the framework of equation
\eqref{DivForm} with $s = -1$, $B = -1$, $G_1 = 4$ and $G_2 = 12$.
Line 1 corresponds to the initial condition at $t = 0$ (blue) and
line 2 is the solution at $t = 1800$ (red). The initial pulse
parameters are $P = 0.275$ and $L = 3.175$.}%
\label{fig:5BC1MS}%
\end{figure}

\begin{figure}
\includegraphics[width = 10cm]{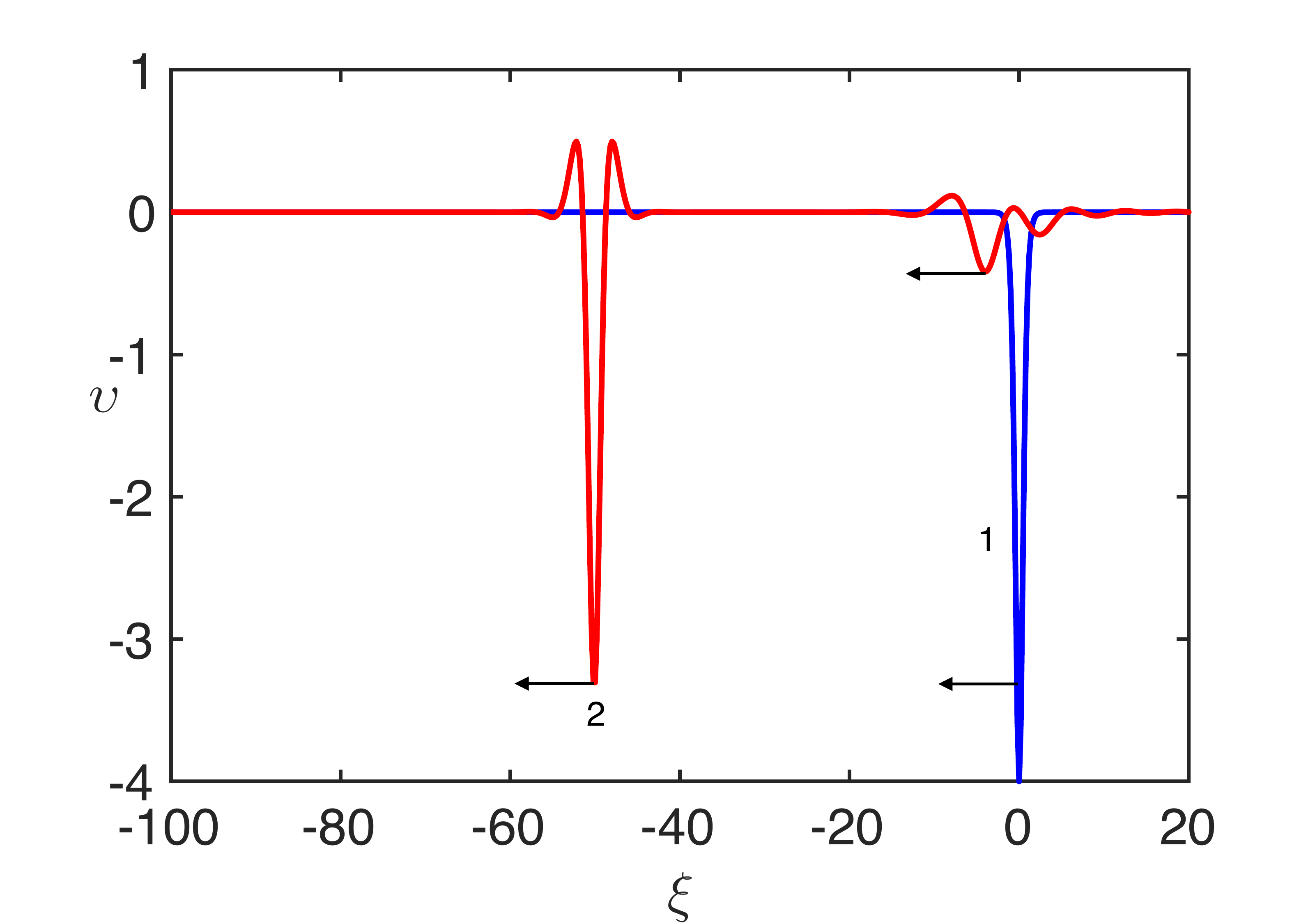}
\caption{(color online) Generation of regular solitons from the
initial sech$^2$-pulse within the framework of equation
\eqref{DivForm} with $s = -1$, $B = 1$, $G_1 = -1.88$ and $G_2 =
-5.64$. Line 1 corresponds to the initial condition at $t = 0$
(blue) and line 2 is the solution at $t = 10$ (red). The initial
pulse parameters are $P = -4$ and $L = 0.3$.}
\label{fig:5BC2MS}%
\end{figure}

Following the same steps as in subsection \ref{Subsec4.1}, we consider the cases when the parameters are (i) $s = -1$, $B = -1$, $G_1 = 4$, and $G_2 = 12$ (see Figure \ref{fig:G23G1sm1Bm1}b) and (ii) $s = -1$, $B = 1$, $G_1 = -1.88$, and $G_2 = -5.64$ (see Figure \ref{fig:G23G1sm1Bp1}b). First of all we observed steady propagation of solitons in both of these cases. Then we observed the emergence of a number of solitons from an initial pulse with larger amplitude and width; this is illustrated in Figure \ref{fig:5BC1MS} for case (i) and Figure \ref{fig:5BC2MS} for case (ii).

\begin{figure}
\includegraphics[width = 10cm]{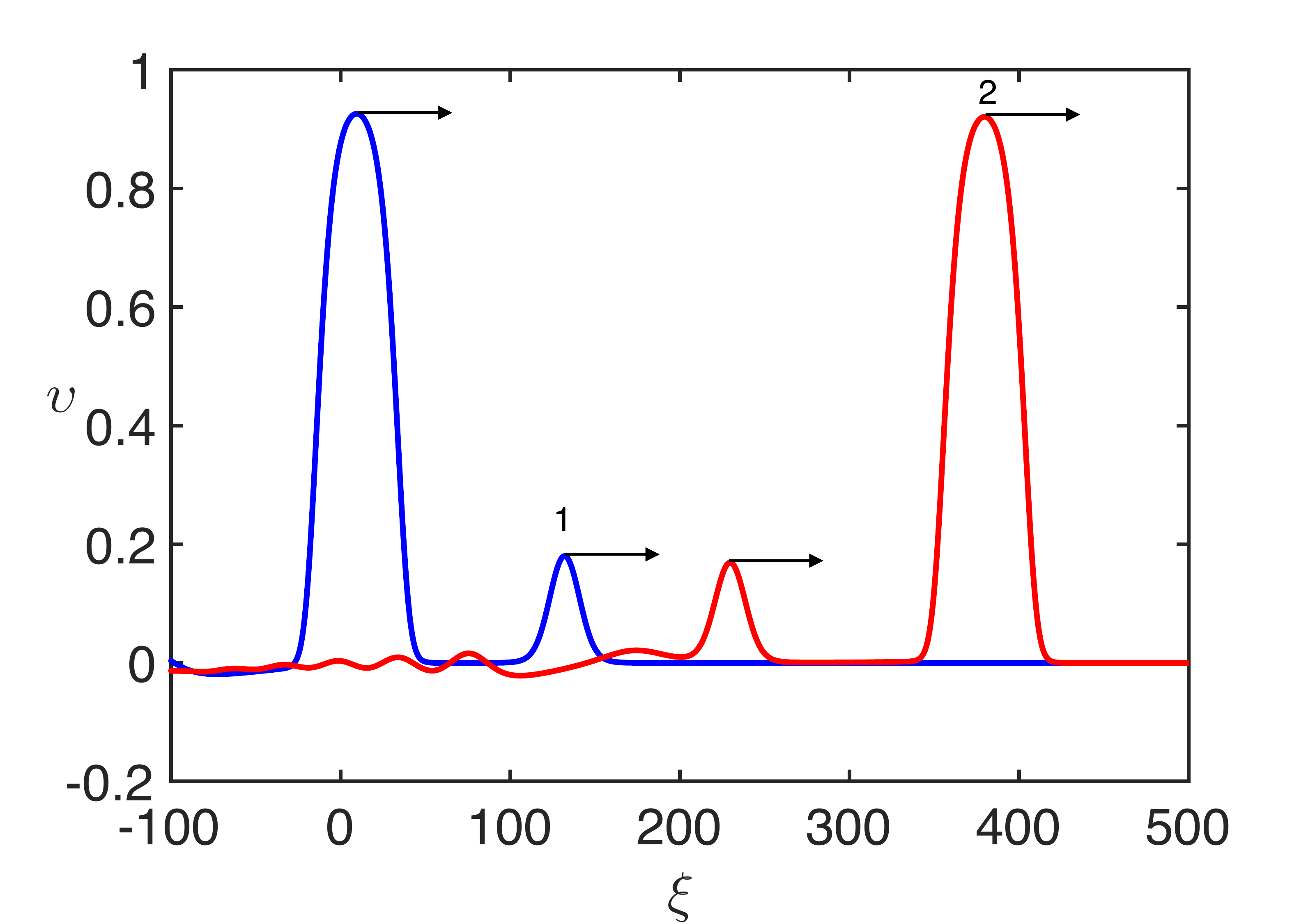}
\caption{(color online) Interaction of two regular solitons within
the framework of equation \eqref{DivForm} with $s = -1$, $B = -1$,
$G_1 = 4$ and $G_2 = 12$. Line 1 corresponds to the initial
condition at $t = 0$ (blue) before the interaction and line 2 to
the solution after the interaction at $t = 2100$ (red). Both
solitons were obtained numerically from a pulse-like initial
condition. The initial pulse parameters are $P_1 = 1.2$ and $L_1 =
5$ for the taller soliton and $P_2 = P_1/3$ and $L_2 = 1$ for the
shorter
soliton.}%
\label{fig:5BC1Col}%
\end{figure}

\begin{figure}
\includegraphics[width = 10cm]{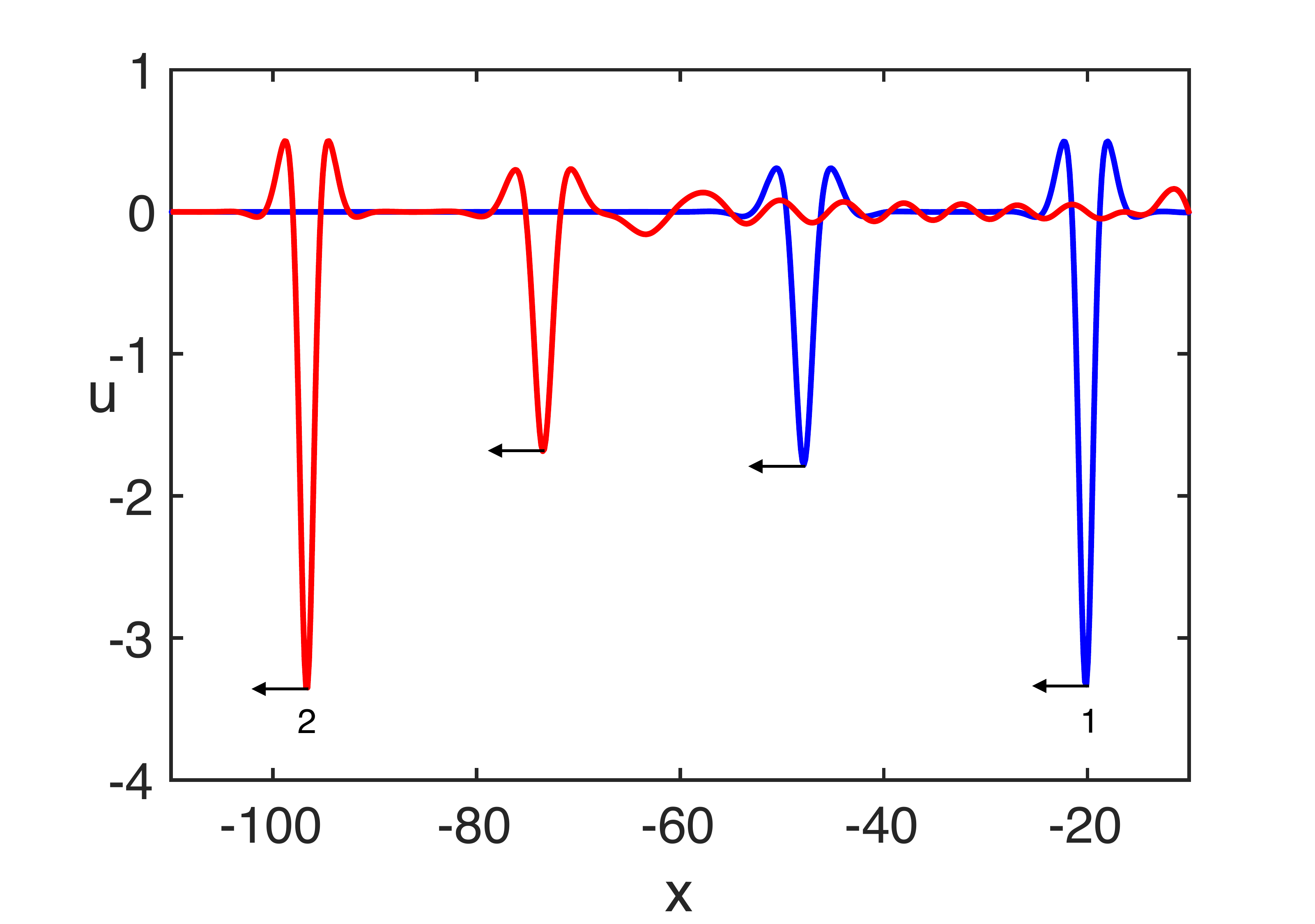}
\caption{(color online) Interaction of two regular solitons within
the framework of equation \eqref{DivForm} with $s = -1$, $B = 1$,
$G_1 = -1.88$ and $G_2 = -5.64$. Line 1 corresponds to the initial
condition at $t = 0$ (blue) before the interaction and line 2 to
the solution after the interaction at $t = 15$ (red). Both
solitons were obtained numerically from a pulse-like initial
condition. The initial pulse parameters are $P_1 = -4$ and $L_1 =
0.3$ for the taller soliton and
$P_2 = P_1/2$ and $L_2 = 0.3$ for the shorter soliton.}%
\label{fig:5BC2Col}%
\end{figure}

Finally we studied the interaction of these solitons. When
colliding two regular solitons, we observed that after the
interaction both the solitons survive, but some portion of their
energy converts into a small wave packet generated in the trailing
wave field; this is shown in Figure \ref{fig:5BC1Col} for case (i)
and Figure \ref{fig:5BC2Col} for case (ii). This is typical for
the inelastic interaction.

\begin{figure}
\includegraphics[width = 10cm]{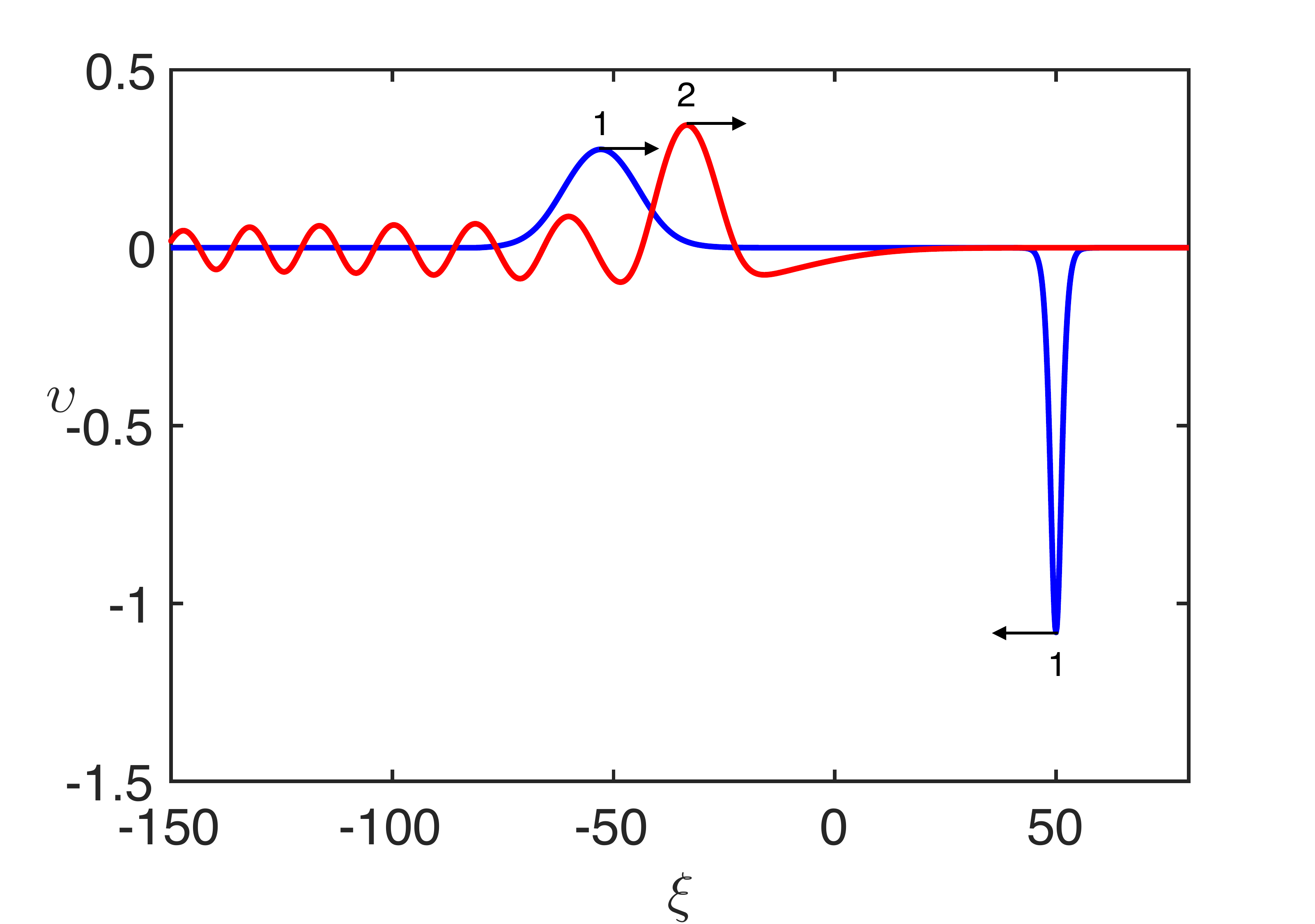}
\caption{(color online) Interaction of regular soliton of positive
polarity and embedded soliton of negative polarity within the
framework of equation \eqref{DivForm} with $s = -1$, $B = -1$,
$G_1 = 4$ and $G_2 = 12$. Line 1 corresponds to the initial
condition at $t = 0$ (blue) before the interaction and line 2 to
the solution after the interaction at $t = 250$ (red). The regular
soliton was obtained numerically from a pulse-like initial
condition. The initial pulse parameters are $P = 0.4$ and $L = 1$.
The gap soliton survives, but the embedded soliton does not.}
\label{fig:5BC1ColES}%
\end{figure}

For the collision of a regular soliton with an embedded soliton,
in case (i) we observed that only the regular soliton survives the
collision and an intense wave packet is generated in the trailing
wave field (see Figure \ref{fig:5BC1ColES}).

\begin{figure}
\includegraphics[width = 10cm]{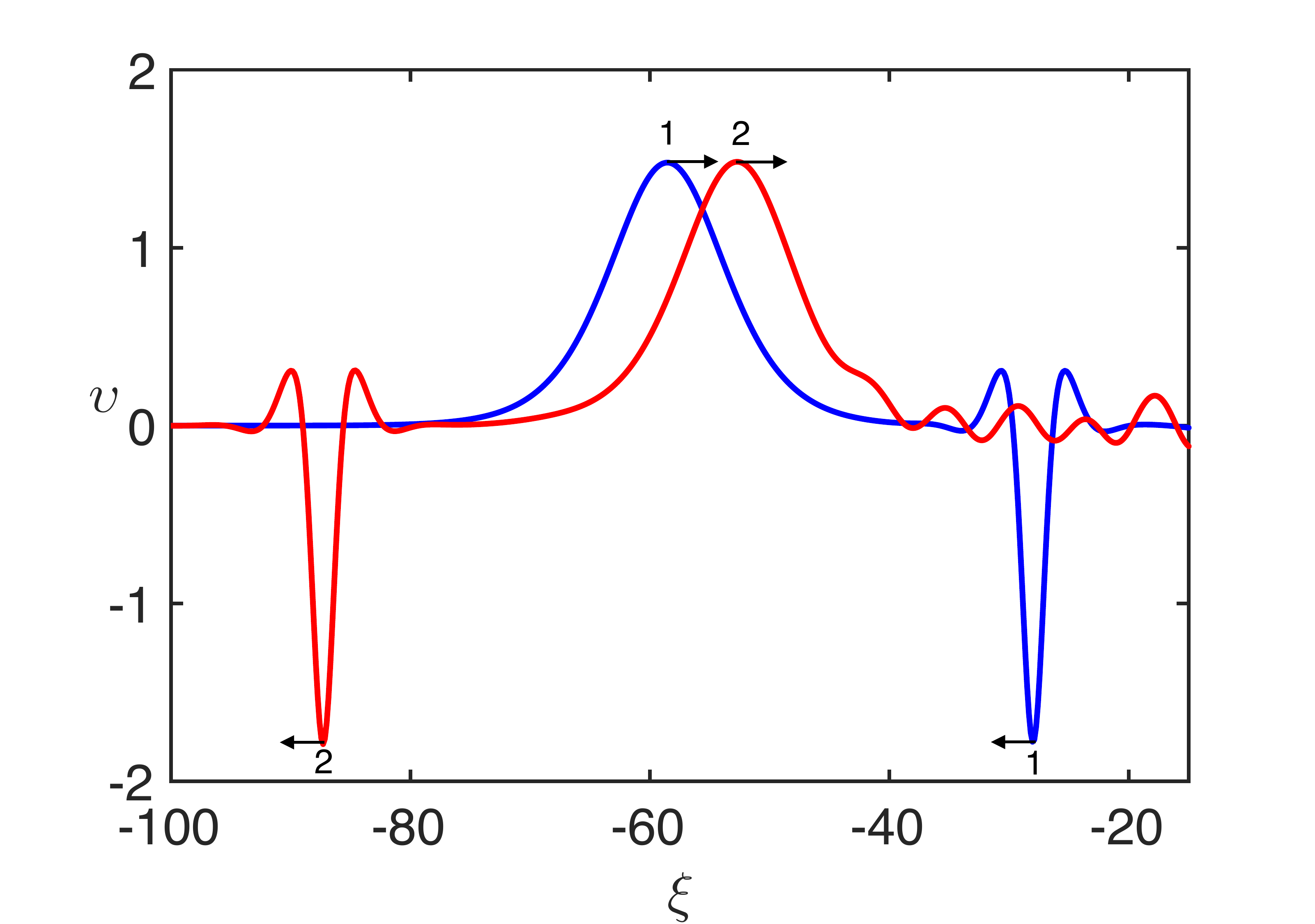}
\caption{(color online) Interaction of regular soliton of negative
polarity and embedded soliton of positive polarity within the
framework of equation \eqref{DivForm} with $s = -1$, $B = 1$, $G_1
= -1.88$ and $G_2 = -5.64$. Line 1 corresponds to the initial
condition at $t = 0$ (blue) before the interaction and line 2 to
the solution after the interaction at $t = 30$ (red). The regular
soliton was obtained numerically from a pulse-like initial
condition. The initial pulse parameters are $P = 0.4$ and $L =
1$.}
\label{fig:5BC2ColES}%
\end{figure}

In contrast to that, in case (ii), the regular and embedded
solitons both survive after the collision and a wave packet is
generated in front of the embedded soliton (see Figure
\ref{fig:5BC2ColES}). In both cases we conclude that the collision
is inelastic. It is worth noting in this case that our numerics
were not stable beyond the time considered in the calculation; it
may be stable for a smaller spatial discretisation, however the
corresponding time discretisation becomes very small and the
calculations take a very long time.

\subsection{The generalised Kawahara equation}
\label{Subsec4.3}%
\begin{figure}[h!]
\includegraphics[width = 10cm]{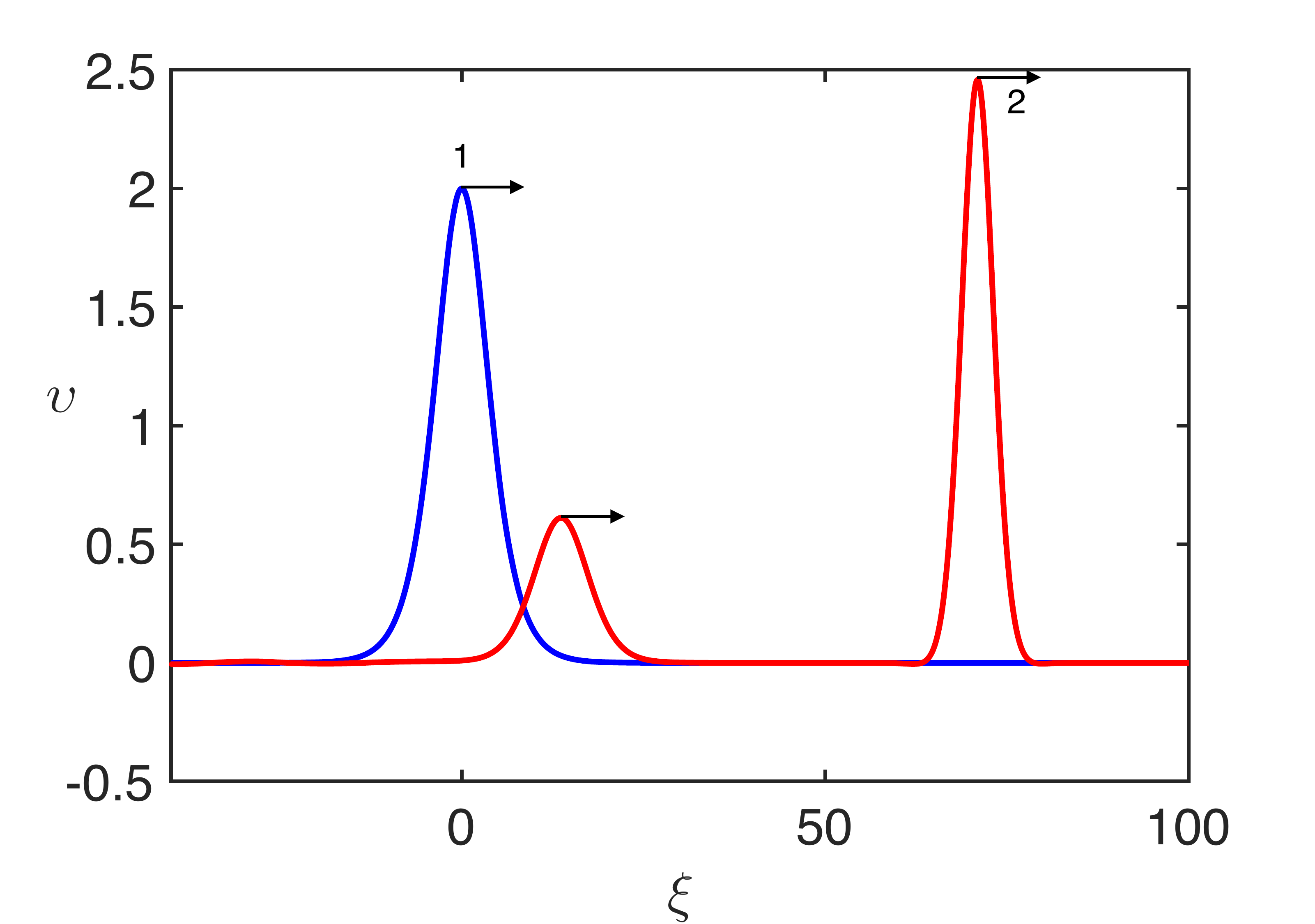}
\caption{(color online) Generation of solitons from the initial
sech$^2$-pulse for the generalised Kawahara equation. Line 1
corresponds to the initial condition at $t = 0$ (blue) and line 2
is the solution at $t = 80$ (red). The initial pulse parameters
are $P = 2$ and $L = 2$.}
\label{Fig:GKMS}%
\end{figure}

As we defined in Section \ref{Subsec3.3} above, the generalised
Kawahara equation is a particular case of equation \eqref{DivForm}
with the coefficients $s = G_1 = G_2 = 0$. We analysed the
stationary solitary solutions, one of which is the YT soliton
\eqref{YT-Soliton}. Here we show that solitary waves emerge from
pulse-type initial perturbations within the generalised Kawahara
equation. As the initial condition the sech$^2$ pulse was chosen.
Figure \ref{Fig:GKMS} illustrates an example of pulse
disintegration onto two solitary waves accompanied by small
residual wave train (not visible in the figure).

\begin{figure}[h!]
\includegraphics[width = 10cm]{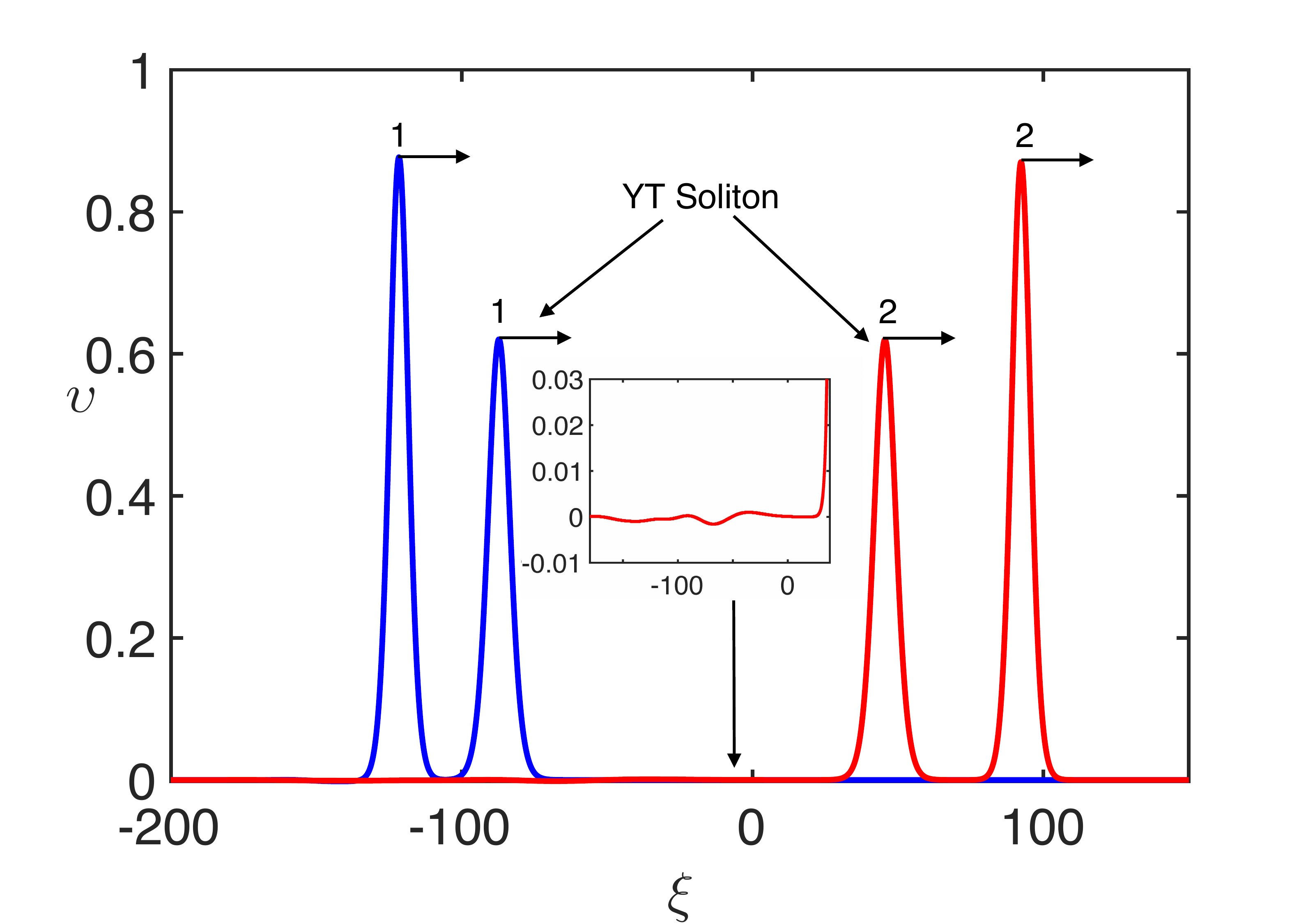}
\caption{(color online) Interaction of the YT regular soliton
(smaller amplitude pulse in the figure) with another regular
solitary wave numerically constructed within the framework of
generalised Kawahara equation. Line 1 corresponds to the initial
condition at $t = 0$ (blue) before the interaction and line 2 to
the solution after the interaction at $t = 680$ (red). The regular
soliton with the framework of the generalised Kawahara equation
was obtained numerically from a pulse-like initial condition. The
initial pulse parameters are $P = 1$ and $L = 1$. Small insertion
shows a magnified fragment of a radiated wave field.}
\label{Fig:GKColES}%
\end{figure}

The soliton interaction is demonstrated in Figure
\ref{Fig:GKColES}. The initial condition was chosen as the YT
soliton (1) and the numerically obtained solitary wave (2). We see
that solitary waves survive the collision and appear after that
with almost the same amplitudes. However, a small wave train
appears in the result of interaction, evidence that the
interaction is inelastic.

\section{Conclusion}
\label{Sect5}%
In this paper we have studied the properties of soliton solutions of the fifth-order KdV equation \eqref{KdV2K} which is used to describe surface and internal gravity waves, as well as appearing in other applied areas. Using the changes of independent and dependent variables this equation has been reduced to the dimensionless form \eqref{DivForm} with the minimum number of independent parameters, only three for positive and negative cubic nonlinearity. In the theory of nonlinear internal waves both these cases of positive and negative cubic nonlinearity can occur depending on the stratification of the fluid \cite{Talipova99, Apel07, RuvKurk14}. In some particular cases equation \eqref{DivForm} reduces to well known equations, among which there are Gardner \cite{OstrEtAl15}, Kawahara \cite{Kawahara72} equations and their generalisation \cite{Giniyatullin14, Kurkina15}, Sawada--Kotera and Kaup--Kupershmidt equations \cite{Dodd82, Newell85, Kichenassamy92}.

Equation \eqref{DivForm} is non-integrable, in general, and it
does not provide conservation of the ``wave energy". However, it
permits the existence of solitary wave solutions both with
monotonic and non-monotonic profiles and even with oscillatory
asymptotics, which suggests the existence of more complicated
structures in the form of stationary multi-solitons
\cite{Gorshkov76, Gorshkov79, Kurkina15}. Following
\cite{Karczewska14}, we have derived an exact soliton solution in
the general case and have shown that in some cases this solution
can represent a regular soliton, whereas in others it represents
an embedded soliton whose speed coincides with the speed of some
linear wave. Our numerical simulations have confirmed that such
solitons can propagate without loss of energy, if there are no
perturbations in the form of other waves or medium inhomogeneity,
or dissipation. We have identified the areas on the plane of
parameters where the derived soliton solution exists as the
embedded or regular soliton and found numerically a number of
soliton solutions of equation \eqref{DivForm} with different sets
of governing parameters.

In the last section we have studied numerically the emergence of
solitons from arbitrary pulse-type initial conditions and have
demonstrated that regular solitons are generated from the initial
conditions. However in the course of interaction such solitons
produce small-amplitude trailing waves which evidences that the
interaction has inelastic character. Moreover, the solitons with
non-monotonic profiles can form stationary or non-stationary bound
states similar to those experimentally observed in
\cite{Gorshkov76}.

One particularly interesting observation emerging from our study
is the stability of embedded solitons with respect to interactions
with other waves (e.g., the regular solitons, like in this paper).
This sheds some additional light on the stability problem and
nature of embedded solitons and indicates that they could be
observed in natural and laboratory environments. In particular, in
the absence of nonlinear dispersion, the embedded soliton survives
the interaction with no visible loss of amplitude. On the other
hand, the results of our numerical investigations suggest that in
some cases embedded solitons, apparently, can transfer to {\it
radiating solitons} under the influence of external perturbations
(the radiating solitons are quasi-stationary long-living solitary
waves permanently radiating from one side small-amplitude linear
waves -- see, e.g., \cite{Grimshaw17, Khusnutdinova17} and
references therein). We do not touch this interesting possibility
in the present paper, but it can be a theme for further study.

Finally, it would be interesting to extend the present study to
the study of the respective ring wave counterparts (see \cite{J,
L, KZ_JFM, KZ_PhysD} and references therein). The  solitary wave
solutions studied in our present paper provide meaningful
``initial conditions" for numerical experimentation with the
amplitude equation describing ring waves.

\acknowledgments The authors are thankful to T. Marchant for
useful discussions and valuable comments. This research was
initiated within the framework of the Scheme 2 grant of the London
Mathematical Society (LMS), in November--December 2015. K.R.K. and
Y.A.S. are grateful to the LMS for the support. Y.A.S.
acknowledges the funding of this study from the State task program
in the sphere of scientific activity of the Ministry of Education
and Science of the Russian Federation (Project No.
5.1246.2017/4.6).  K.R.K. acknowledges the support received during
her stay at ESI, Vienna in 2017, where some parts of this work
were finalised. M.R.T acknowledges the support of the
Engineering and Physical Sciences Research Council (EPSRC).\\

\appendix
\section{Coefficients of the Higher-Order KdV Equation for Water Waves}
\label{AppendA}%

The coefficients of equation \eqref{KdV2K} for gravity surface waves are given in \eqref{A9}. Here we present the coefficients for internal waves in two-layer fluid derived in \cite{Giniyatullin14}. All notations are shown in Figure \ref{fig:Sketch}.

\begin{align}
c &= \sqrt{\frac{(\rho_2 - \rho_1) gh_1h_2}{\rho_1h_2 + \rho_2h_1}}, \quad \alpha = \frac{3c}{2h_1h_2} \frac{\rho_2h_1^2 - \rho_1h^2_2}{\rho_1h_2 + \rho_2h_1}, \quad \beta = \frac{h_1h_2}{6} c \frac{\rho_1h_1 + \rho_2h_2 - 3 \sigma/c^2}{\rho_1h_2 + \rho_2h_1}, \label{A1} \\
\alpha_1 &= -\frac{3c}{8h_1^2h_2^2} \frac{\rho_2^2h_1^4 + \rho_1^2h_2^4 + 2\rho_1\rho_2h_1h_2\lb 4h_1^2 + 7h_1h_2 + 4h_2^2\rb}{\lb \rho_1h_2 + \rho_2h_1\rb^2}, \label{A2} \\
\beta_1 &= \frac{ch_1h_2}{90} \frac{\rho_1 \rho_2 \lb h_1^4 + h_2^4 + 15h_1^2h_2^2/2 \rb + \lb 19/4 \rb h_1h_2 \lb \rho_1^2h_1^2 + \rho_2^2h_2^2\rb - S_1}{\lb \rho_1h_2 + \rho_2h_1\rb^2}, \label{A3} \\
\gamma_1 &= \frac{c}{12}\frac{5h_1h_2 \lb \rho_2^2h_1 - \rho_1^2h_2 \rb + \rho_1 \rho_2 \lb h_1 - h_2 \rb \lb 7h_1^2 + 9h_1h_2 + 7h_2^2 \rb - 2S_2}{\lb \rho_1h_2 +\rho_2h_1\rb^2}, \label{A4} \\
\gamma_2 &= \frac{c}{24}\frac{23h_1h_2 \lb \rho_2^2h_1 - \rho_1^2h_2 \rb + \rho_1\rho_2 \lb h_1 - h_2 \rb \lb 31h_1^2 + 39h_1h_2 + 31h_2^2 \rb + 2S_2/5}{\lb \rho_1h_2 + \rho_2h_1\rb^2}, \label{A5}
\end{align}
where
\begin{equation*}
S_1 = 5h_1h_2\frac{3\sigma}{2c^2}\lb \rho_1h_1 + \rho_2h_2 + \frac{3\sigma}{2c^2}\rb, \quad S_2 = \frac{3\sigma}{2c^2}\lb \rho_2h_1^2 - \rho_1h_2^2\rb.
\end{equation*}

In particular, when $\rho_1 = 0$, we obtain the coefficients of
equation \eqref{KdV2K} for surface gravity-capillary waves on a
thin liquid layer:
\begin{align}
c = \sqrt{gh_2}, \quad \alpha = \frac{3c}{2h_2},  \quad \alpha_1 = -\frac{3c}{8h_2^2}, \quad \beta = \frac{ch_2^2}{6} \lb 1 - \frac{3 \sigma}{c^2 \rho_2h_2} \rb, \label{A6} \\
\beta_1 = \frac{ch_2^4}{18} \lsq \frac{19}{20} - \frac{3\sigma}{2c^2 \rho_2h_2} \lb 1 + \frac{3 \sigma}{2c^2 \rho_2h_2} \rb \rsq, \label{A7} \\
\gamma_1 = \frac{5ch_2}{12} \lb 1 - \frac{3 \sigma}{5c^2 \rho_2h_2} \rb, \quad \gamma_2 = \frac{23ch_2}{24} \lb 1 + \frac{15 \sigma}{23c^2 \rho_2h_2} \rb. \label{A8}
\end{align}

\section{Petviashvili's Method}
\label{AppendB}%

Let us make a Fourier transform of equation \eqref{SolODE} with respect to the variable $\zeta$ denoting the Fourier image of the function $\upsilon(\zeta)$ by $\hat F(\upsilon)$:
\begin{equation}
\lb B \kappa^4 - \kappa^2 - V \rb \hat{F}(\upsilon) = -\frac{s}{3} \hat{F}(\upsilon^3) - \frac 12 \lb 1 - G_1 \kappa^2 \rb \hat{F}(\upsilon^2) - \frac 12 \lb G_2 - 3G_1 \rb \hat{F} \lsq \lb\upsilon'\rb^2 \rsq,
\label{B1}
\end{equation}
where $\kappa$ is the parameter of the Fourier transform (the dimensionless wave number).

If we multiply equation \eqref{B1} by $\hat{F}(\upsilon)$ and integrate it with respect to $\kappa$ from minus to plus infinity, we obtain the equality
\begin{equation}
\int\limits_{-\infty}^{+\infty} \lb B \kappa^4 - \kappa^2 - V \rb \lsq \hat{F}(\upsilon) \rsq^2\,\mathrm{d}\kappa = -\frac 12\int\limits_{-\infty}^{+\infty} \lset \frac{2s}{3} \hat{F}(\upsilon^3) + \lb 1 - G_1\kappa^2 \rb \hat{F}(\upsilon^2) + \lb G_2 - 3G_1 \rb \hat{F} \lsq \lb \upsilon' \rb^2 \rsq \rset \hat{F}(\upsilon)\,\mathrm{d}\kappa.
\label{B2}
\end{equation}
If $\upsilon(\zeta)$ is an exact solution of equation \eqref{SolODE} and $\hat{F}(\upsilon)$ is its Fourier image satisfying equation \eqref{B1}, then it follows from equation \eqref{B2} that the quantity $M$, dubbed the stabilising factor and defined below, should be equal to one:
\begin{equation}
M [\upsilon] = \frac{-2 \int \limits_{-\infty}^{+\infty} \lb B \kappa^4 - \kappa^2 - V \rb \lsq \hat{F}(\upsilon) \rsq^2\, \mathrm{d}\kappa}
{\int\limits_{-\infty}^{+\infty} \lset (2s/3) \hat{F}(\upsilon^3) + \lb 1 - G_1 \kappa^2 \rb \hat{F}(\upsilon^2) + \lb G_2 - 3G_1 \rb \hat{F} \lsq \lb \upsilon' \rb^2 \rsq \rset \hat{F}(\upsilon)\,\mathrm{d} \kappa}.
\label{B3} %
\end{equation}
However, in general, if $\upsilon(\zeta)$ is not a solution of equation \eqref{SolODE}, then $M[\upsilon]$ is some functional of $\upsilon$. In the spirit of the Petviashvili method, let us construct the iteration scheme (for details see \cite{Petviashvili92, Pelinovsky04}):
\begin{equation}
\hat{F}(\upsilon_{n+1}) = -\frac{1}{2} M^r [\upsilon_n] \frac{(2s/3) \hat{F}(\upsilon_n^3) + \lb 1 - G_1 \kappa^2 \rb \hat{F}(\upsilon_n^2) + \lb G_2 - 3G_1 \rb \hat{F} \lsq \lb\upsilon_n'\rb^2 \rsq}{B \kappa^4 - \kappa^2 - V},
\label{B4}
\end{equation}
where the factor $M$ is used to provide a convergence of the iterative scheme (otherwise the scheme is not converging), and the exponent $r$ should be taken in the range $r = [3/2, 2]$. As has been shown in \cite{Pelinovsky04}, $r = 3/2$ provides the fastest convergence for pure cubic nonlinearity, whereas $r = 2$ provides the fastest convergence for pure quadratic nonlinearity. In our calculations we chose $r = 7/4$ which provided the fastest convergence to the stationary solution for mixed quadratic and cubic nonlinearity.

The convergence is controlled by the closeness of the parameter $M$ to unity. Starting from the arbitrary pulse-type function $\upsilon_0(\zeta)$, we conducted calculations with the given parameters $B$, $G1$, $G2$, and $V$ on the basis of the iteration scheme \eqref{B4} until the parameter $M$ was close to 1, up to small quantity $\epsilon$, i.e., until $|M - 1| \le \epsilon$ (in our calculations it was set to $\epsilon = 10^{-6}$).

To avoid a singularity in equation \eqref{B4}, the speed of a
solitary wave should be chosen in such a way that the
fourth-degree polynomial in the denominator of equation \eqref{B4}
does not have real roots. This corresponds to the case when there
is no resonance between the solitary wave and a linear wave, i.e.,
$V \ne V_{ph}(\kappa)$ -- see equation \eqref{DispRel}. Under this
condition only regular solitons can be constructed by means of
this method, not the embedded soliton.

\section{Pseudospectral Scheme for the Fifth-Order KdV Equation}
\label{AppendC}%

The numerical scheme used for the interaction of regular and
embedded solitons is as follows. We implement a pseudospectral
scheme using a \nth{4} order Runge--Kutta method for time
stepping. The time stepping is performed in the Fourier space and
the nonlinear terms are calculated in the real space and
transformed back to the Fourier space for use in the method.
Firstly we consider a solution in the domain $[-L, L]$ and
transform it to the domain $[0, 2\pi]$ via the transform
$\tilde{x} = S x  + \pi$ with $S = \pi/L$. Writing \eqref{KdV2K}
in the divergent form, we obtain (omitting tildes)
\begin{equation}
u_{t} + S \pdiff{ }{x} \left [\alpha \frac{u^2}{2} + \alpha_1 \frac{u^3}{3} + \beta S^2 \pdiffn{2}{u}{x} + \beta_1 S^4 \pdiffn{4}{u}{x} + \frac{\gamma_1}{2} S^2 \pdiffn{2}{u^2}{x} + \frac{\gamma_2 - 3\gamma_1}{2} \left (S \pdiff{u}{x} \right )^2 \right]= 0.
\label{KdV2K2Pi}
\end{equation}
The terms $u^2$, $u^3$ and $\left(\pdiff{u}{x}\right)^2$ are calculated in the real domain before transforming back to the Fourier space for use in the time-stepping algorithm.

Let us discretise the solution interval by $N$ nodes where $N$ is a power of 2, so we have spacing $\Delta x = 2 \pi/N$ (in our calculation we used $N = 2^{13} = 8192$, so that the spacial resolution was $\Delta \xi = 0.15$). We use the Discrete Fourier Transform (DFT)
\begin{equation}
\hat{u} (k, t) = \frac{1}{\sqrt{N}} \sum_{j=0}^{N-1} u (x_j, t) e^{-ikx_j}, \quad -\frac{N}{2} \leq k \leq \frac{N}{2} - 1,
\label{DFT}
\end{equation}
where $x_j = j \Delta x$, and $k$ is an integer representing the discretised (and scaled) wavenumber. The inverse transform is
\begin{equation}
u (x, t) = \frac{1}{\sqrt{N}} \sum_{k=-N/2}^{N/2-1} \hat{u} (k, t) e^{ikx_j}, \quad j=0, 1, \dots, N-1.
\label{IDFT}
\end{equation}
We make use of the Fast Fourier Transform (FFT) algorithm to implement these transforms effectively. We introduce the following notation for the last term in square brackets of equation \eqref{KdV2K2Pi} to simplify the expression: $\hat{z} = \mathscr{F} \lb \mathscr{F}^{-1} \lb i k S \hat{u} \rb^2 \rb$, where $\mathscr{F}$ and $\mathscr{F}^{-1}$ represent the forward and inverse Fourier transforms respectively. Applying these transforms to \eqref{KdV2K2Pi} we obtain
\begin{equation}
\hat{u}_{t} = F(\hat{u}) \equiv -\lsq \frac{\alpha i k S}{2}\widehat{u^2} + \frac{\alpha_1 i k S}{3} \widehat{u^3} - \beta i k^3 S^3 \hat{u} + \beta_1 i k^5 S^5 \hat{u} - \frac{\gamma_1}{2} i k^3 S^3 \widehat{u^2} + \frac{\gamma_2 - 3\gamma1}{2} i k S \hat{z} \rsq.
\label{KdVFS}
\end{equation}
To solve the ODE \eqref{KdVFS} numerically, we use a \nth{4} order Runge--Kutta method for time stepping. Let us assume that the solution at time $t$ is given by $\hat{u}_j = \hat{u} (x, j \Delta t)$, where $\Delta t$ is the time step of integration. The solution at time $t = (j+1)\Delta t$ is given by
\begin{equation}
\hat{u}_{j+1} = \hat{u}_{j} + \frac{1}{6} \lb a_{j} + 2 b_{j} + 2 c_{j} + d_{j} \rb,
\end{equation}
where complex quantities $a$, $b$, $c$, and $d$ are defined as
\begin{align}
&a_{j} = \Delta t F \lb \hat{u}_{j} \rb, &&b_{j} = \Delta t F \lb \hat{u}_{j} + \frac{1}{2} a_{j} \rb, \notag \\
&c_{j} =  \Delta t F \lb \hat{u}_{j} + \frac{1}{2} b_{j} \rb, &&d_{j} = \Delta t F \lb \hat{u}_{j} + c_{j} \rb.
\end{align}
The nonlinear terms $u^2$ and $u^3$ were evaluated in the real space and then were transformed to the Fourier space for use in equation \eqref{KdVFS}.

Due to periodic boundary conditions (which is the intrinsic
feature of the pseudospectral method), the radiated waves can
re-enter the region of interest and interfere with the main wave
structures. To alleviate this, we have introduced a damping region
(``sponge layer'') at each end of the domain to prevent waves
re-entering. Within the sponge layer we introduce in the left-hand
side of equation \eqref{KdV2K2Pi} a linear decay term $\nu\:
r(x)\: u$, where $\nu$ is the coefficient of artificial viscosity
and
\begin{equation}%
\label{Sponge} %
r(x) = \frac{1}{2} \lsq 2 +  \tanh D \lb x - \frac{3L}{4} \rb - \tanh D \lb x + \frac{3L}{4} \rb \rsq.
\end{equation}
The coefficient $D$ was chosen such that damping occurs only beyond the region of interest and does not affect the main wave structures. The spatially nonuniform decay term was treated numerically in the same way as the nonlinear terms.

\end{document}